\documentclass[a4paper]{article}
\usepackage{a4wide}
\usepackage{graphicx}
\usepackage{amsmath}
\usepackage{amsthm}
\usepackage{amssymb, latexsym, mathrsfs}
\usepackage{dsfont}
\usepackage{enumerate}
\usepackage{subfigure}
\usepackage{algorithm}
\usepackage{color}
\usepackage{transparent}
\usepackage{url}
\usepackage[affil-it]{authblk}
\theoremstyle{plain}
\newtheorem{thm}{Theorem}
\newtheorem{assum}{Assumption}

\newtheorem{prop}{Proposition}
\newtheorem{lem}{Lemma}

\def\Rset{\mathbb{R}}
\def\Nset{\mathbb{N}}
\def\Zset{\mathbb{Z}}

\newcommand{\BB}{{\mathcal B}}
\newcommand{\CC}{{\mathcal C}}

\newcommand{\FF}{{\mathcal F}}

\newcommand{\HH}{{\mathcal H}}

\newcommand{\KK}{{\mathcal K}}

\newcommand{\MM}{{\mathcal M}}
\newcommand{\NN}{{\mathcal N}}

\newcommand{\PP}{{\mathcal P}}

\newcommand{\SSS}{{\mathcal S}}

\newcommand{\hx}{{\hat x}}

\newcommand{\bx}{{\bar x}}

\newcommand{\bv}{{\bar v}}

\newcommand{\mbf}[1]{\mathbf{#1}}
\newcommand{\px}{{x^+}}
\newcommand{\subss}[2]{#1_{[#2]}}
\newcommand{\tx}{{\tilde x}}

\newcommand{\hpx}{{\hat{x}}^+}

\newcommand{\pz}{{{z}^+}}
\newcommand{\diag}{\mbox{diag}}
\newcommand{\rank}{\mbox{rank}}
\newcommand{\Xset}{\mathbb{X}}
\newcommand{\bXset}{{\bar{\mathbb{X}}}}

\newcommand{\Uset}{\mathbb{U}}

\newcommand{\hXset}{\mathbb{{\hat{X}}}}

\newcommand{\Wset}{\mathbb{{{W}}}}

\newcommand{\Vset}{\mathbb{V}}
\newcommand{\ball}[1]{{B_{#1}}}

\newcommand{\norme}[2]{{||{#1}||_{#2}}}

\newcommand{\hXi}{\hat{\Xi}}
\newcommand{\hd}{\hat{d}}
\newcommand{\hl}{\hat{l}}
\newcommand{\hf}{\hat{f}}
\newcommand{\hr}{\hat{r}}

\newcommand{\Pset}{\mathbb{P}}
\newcommand{\hFF}{{\hat{\mathcal F}}}
\newcommand{\One}{\textbf{1}}
\newcommand{\Zero}{\textbf{0}}
\newcommand{\Metz}{\mathds{M}}

\newcommand{\eye}[1]{\mathds{I}_#1}
\newcommand{\hbx}{{\bar{\hat x}}}

\newcommand{\oneblock}[1]{
\begin{aligned}
#1
\end{aligned}
}

\newcommand{\imply}{\Rightarrow}

\newcommand{\ba}[1]{\begin{array}{#1}}
\newcommand{\ea}{\end{array}}

\newcommand{\matr}[1]{
\begin{bmatrix}
    #1
\end{bmatrix}
}

\begin{document}

     \title{Plug-and-Play Decentralized Model Predictive Control \thanks{The research leading to these results has received funding from the European Union Seventh Framework Programme [FP7/2007-2013]  under grant agreement n$^\circ$ 257462 HYCON2 Network of excellence.}}
     \author{Stefano Riverso%
       \thanks{Electronic address: \texttt{stefano.riverso@unipv.it}; Corresponding author}} 
     
     \author{Marcello Farina%
       \thanks{Electronic address: \texttt{farina@elet.polimi.it}}} 

     \author{Giancarlo Ferrari-Trecate%
       \thanks{Electronic address: \texttt{giancarlo.ferrari@unipv.it}\\S. Riverso and G. Ferrari-Trecate are with the Dipartimento di Informatica e Sistemistica, Universit\`a degli Studi di Pavia, via Ferrata 1, 27100 Pavia, Italy\\M. Farina is with the Dipartimento di Elettronica e Informazione, Politecnico di Milano, via Ponzio 34/5, 20133 Milan, Italy}} 
     \affil{Dipartimento di Informatica e Sistemistica\\Universit\`a degli Studi di Pavia}

     \date{\textbf{Technical Report}: RIDIS 157/12\\ March, 2012}

     \maketitle

     \begin{abstract}
       In this paper we consider a linear system structured into physically coupled subsystems and propose a decentralized control scheme capable to guarantee asymptotic stability and satisfaction of constraints on system inputs and states. The design procedure is totally decentralized, since the synthesis of a local controller uses only information on a subsystem and its neighbors, i.e. subsystems coupled to it. We first derive tests for checking if a subsystem can be plugged into (or unplugged from) an existing plant without spoiling overall stability and constraint satisfaction. When this is possible, we show how to automatize the design of local controllers so that it can be carried out in parallel by smart actuators equipped with computational resources and capable to exchange information with neighboring subsystems. In particular, local controllers exploit tube-based Model Predictive Control (MPC) in order to guarantee robustness with respect to physical coupling among subsystems. Finally, an application of the proposed control design procedure to frequency control in power networks is presented.   
     \end{abstract}

     \newpage

     \section{Introduction}
          Decentralized regulators have been studied since the 70's as a viable solution to the control of large-scale systems composed by several physically coupled subsystems \cite{Bakule1988,Lunze1992}. Compared to centralized schemes, decentralized control structures offer several advantages such as parallel computation of control variables, local transmission of information between each subsystem and the corresponding regulator and higher reliability in presence of controller faults. The problem of guaranteeing stability and suitable performance levels for decentralized control systems has been addressed in the 70's and 80's mainly for unconstrained systems \cite{Wang1973,Lunze1992}. Similar remarks apply to distributed control (also known as overlapping decentralized control), where controllers can exchange pieces of information through a communication network (see, e.g. \cite{Lavaei2008}, and references therein).

          Decentralized and distributed control schemes for constrained systems have been proposed only much more recently in the context of Model Predictive Control (MPC) \cite{Camponogara2002,Keviczky2006,Raimondo2007a,Farina2011b,Farina2012,Riverso2012e,Rawlings2009}. These results are particularly appealing because replace large-scale optimization problems stemming from centralized MPC with several smaller-scale problems that can be solved in parallel using computational resources collocated with sensors. While the main focus of decentralized and distributed control is on limiting the computational burden and communication cost associated to real-time coperations of the control system, attention has also been paid to the complexity of the controller design procedure. In this respect, decentralized and distributed controllers can be designed either in a centralized fashion, i.e. relying on the knowledge of the collective model, or in a decentralized fashion, i.e. not requiring the knowledge of the collective model \cite{Bakule1988,Lunze1992}. However, decentralized design does not prevent from using collective quantities, based on pieces of information from all subsystems. An example are decentralized control schemes that rely on vector Lyapunov functions for assessing the stability of the closed-loop system \cite{Lunze1992} and hence require stability analysis of an $M$-th order system where $M$ is the number of subsystems.

          In this paper we move one step further and propose decentralized MPC (DeMPC) schemes with Plug-and-Play (P\&P) capabilities. Similarly to \cite{Stoustrup2009}, P\&P means that
          \begin{enumerate}[(i)]
          \item\label{enu:whypp}the design of a single controller involves at most information about the subsystem under control and its neighbors, i.e. no step of the design procedure involves collective quantities;
          \item when a subsystem joins/leaves an existing plant there is a procedure for
            \begin{enumerate}[(a)]
            \item\label{enu:whyppa} assessing if the operation does not spoil stability and constraint satisfaction for the overall plant;
            \item automatically retuning at most the controllers of the subsystem and its successors, i.e. subsystems influenced by it.
            \end{enumerate}
          \end{enumerate}

          P\&P controllers are attractive for the following reasons. First, the complexity of designing a controller for a given subsystem scales with the number of neighboring subsystem only. Second, P\&P eases the revamping of control systems by enabling the replacement of actuators with limited interaction of human operators. It is well known that, for general interconnection topologies, requirement (\ref{enu:whypp}) above implies the design of regulators for each subsystem that are robust to the coupling with neighboring subsystems \cite{Lunze1992}. Our design procedure is no exception and we will exploit tube-based MPC \cite{Mayne2005} for the design of robust local controllers. While this introduces an unavoidable degree of conservatism, we argue that P\&P DeMPC can be successfully applied in a number of real world plants where coupling among subsystems is sufficiently weak. As an example, we will use P\&P DeMPC for designing the Automatic Generation Control (AGC) layer for frequency control in a realistic power network and discuss the plugging in and unplugging of generators areas.

          The paper is structured as follows. The design of decentralized controllers is introduced in Section \ref{sec:modelcontroller} with a focus on the assumptions needed for guaranteeing asymptotic stability of the origin and constraint satisfaction. In Section \ref{sec:main} we discuss how to design the local controllers in a distributed fashion and in Section \ref{sec:plugplay} we describe P\&P operations. In Section \ref{sec:computational} we discuss the practical design of the local controllers. In Section \ref{sec:example} we present the application of P\&P DeMPC to frequency control in a power network and Section \ref{sec:conclusions} is devoted to concluding remarks. 

          \textbf{Notation.} We use $a:b$ for the set of integers $\{a,a+1,\ldots,b\}$. The column vector with $s$ components $v_1,\dots,v_s$ is $\mbf v=(v_1,\dots,v_s)$. The function $\diag(G_1,\ldots,G_s)$ denotes the block-diagonal matrix composed by $s$ block $G_i$, $i\in 1:s$.  The pseudo-inverse of a matrix $A\in\Rset^{m\times n}$ is denoted with $A^\flat$. The symbol $\oplus$ denotes the Minkowski sum of sets, i.e. $A=B\oplus C$ if and only if $A=\{a:a=b+c,\mbox{ for all }b\in B \mbox{ and }c\in C\}$. Moreover, $\bigoplus_{i=1}^sG_i=G_1\oplus\ldots\oplus G_s$. The symbols $\One_n$ and $\Zero_n$ denote the column vectors with $n$ elements equal to $1$ and $0$, respectively. A zonotope is a centrally symmetric convex polytopes: given a vector $p\in\Rset^n$ and a matrix $\Xi\in\Rset^{n\times m}$, the zonotope $\Xset\subseteq\Rset^n$ is the set $\Xset=\{x~|~x=p+\Xi d,~\norme{d}{\infty}\leq 1\}$, with $d\in\Rset^m$. Moreover, if $\Xset$ is a zonotope, its support function in the direction $c\in\Rset^n$ is given by \cite{Kolmanovsky1998} as 
          \begin{equation}
            \label{eq:suppfunc}
            \sup_{\substack{x\in\Xset}}~c^Tx=\norme{\Xi^Tc}{1}.
          \end{equation}
          The set $\Xset\subseteq\Rset^n$ is Robust Positively Invariant (RPI) \cite{Rawlings2009} for $x(t+1)=f(x(t),w(t))$, $w(t)\in\Wset\subseteq\Rset^m$ if $x(t)\in\Xset\imply f(x(t),w(t))\in\Xset\mbox{, }\forall w(t)\in\Wset$. The RPI set $\underline\Xset$ is minimal if every other RPI $\Xset$ verifies $\underline{\Xset}\subseteq\Xset$. The RPI set $\Xset(\delta)$ is a $\delta$-outer approximation of the minimal RPI $\underline\Xset$ if
          \begin{equation}
            \label{eq:defapproxmRPI}
            x\in\Xset(\delta)\imply\exists~\underline x\in\underline\Xset\mbox{ and } \sigma\in\ball{\delta}(0): x = \underline x + \sigma.
          \end{equation}
where $\ball{\delta}(v)$ is the 2-norm open ball of radius $\delta$ centered in $v\in\Rset^n$.

     \section{Decentralized tube-based MPC of linear systems}
          \label{sec:modelcontroller}
          We consider a discrete-time Linear Time-Invariant (LTI) system
          \begin{equation}
            \label{eq:system}
            \mbf{\px}=\mbf{A}\mbf{x}+\mbf{B}\mbf{u}
          \end{equation}
          where $\mbf x\in\Rset^n$ and $\mbf u\in\Rset^m$ are the state and the input, respectively, at time $t$ and $\mbf{\px}$ stands for $\mbf{x}$ at time $t+1$. We will use the notation $\mbf x(t)$, $\mbf u(t)$ only when necessary. The state is partitioned into $M$ state vectors $\subss x i\in\Rset^{n_i}$, $i\in\MM=1:M$ such that $\mbf{x}=(\subss x 1,\dots,\subss x M)$, and $n=\sum_{i\in\MM}n_i$. Similarly, the input is partitioned into $M$ vectors $\subss u i\in\Rset^{m_i}$, $i\in\MM$ such that $\mbf{u}=(\subss u 1,\dots,\subss u M)$ and $m=\sum_{i\in\MM}m_i$. 

          We assume the dynamics of the $i-th$ subsystem is given by
          \begin{equation}
            \label{eq:subsystem}
            \subss\Sigma i:\quad\subss \px i=A_{ii}\subss x i+B_i\subss u i+\subss w i
          \end{equation}
          \begin{equation}
            \label{eq:couplingW}
            \subss w i = \sum_{j\in\NN_i}A_{ij}\subss x j
          \end{equation}
          where $A_{ij}\in\Rset^{n_i\times n_j}$, $i,j\in\MM$, $B_i\in\Rset^{n_i\times m_i}$ and $\NN_i$ is the set of neighbors to subsystem $i$ defined as $\NN_i=\{j\in\MM:A_{ij}\neq 0, i\neq j\}$. 

          According to \eqref{eq:subsystem}, the matrix $\mbf A$ in \eqref{eq:system} is decomposed into blocks $A_{ij}$, $i,j\in\MM$. We also define $\mbf{A_D}=\diag(A_{11},\dots,A_{MM})$ and $\mbf{A_C}=\mbf{A}-\mbf{A_D}$, i.e. $\mbf{A_D}$ collects the state transition matrices of every subsystem and $\mbf{A_C}$ collects coupling terms between subsystems. From \eqref{eq:subsystem} one also obtains $\mbf{B}=\diag(B_1,\dots,B_M)$ because submodels \eqref{eq:subsystem} are input decoupled.

          In this Section we propose a decentralized controller for \eqref{eq:system} guaranteeing asymptotic stability of the origin of the closed-loop system and constraints satisfaction.
       
          In the spirit of tube-based control \cite{Mayne2005}, we treat $\subss w i$ as a disturbance and equip \eqref{eq:subsystem} with the controller $\subss\CC i$ given by
          \begin{equation}
            \label{eq:tubecontrol}
            \subss u i=\subss v i+K_i(\subss x i-\subss\bx i).
          \end{equation}
          where $K_i\in\Rset^{m_i\times n_i}$, $i\in\MM$ and variables $\subss v i$ and $\subss\bx i$ will be computed by a local state-feedback MPC controller, i.e. there exist functions $\kappa_i:\Rset^{n_i}\rightarrow\Rset^{m_i}$ and  $\eta_i:\Rset^{n_i}\rightarrow\Rset^{n_i}$ such that $\subss v i=\kappa_i(\subss x i)$ and $\subss\bx i=\eta_i(\subss x i)$. Note that the controller $\subss\CC i$ is completely decentralized, since it depends upon quantities of system $\subss\Sigma i$ only.

          Next, we clarify properties of matrices $K_i$, $i\in\MM$ that are required for the stability of system \eqref{eq:system} controlled by \eqref{eq:tubecontrol}. Defining the collective variables $\mbf \bx = (\subss \bx 1,\ldots,\subss \bx M)\in\Rset^n$, $\mbf v = (\subss v 1,\ldots,\subss v M)\in\Rset^m$ and the matrix $\mbf K = \diag(K_1,\ldots,K_M)\in\Rset^{m\times n}$, from \eqref{eq:subsystem} and \eqref{eq:tubecontrol} one obtains the collective model
          \begin{equation}
            \label{eq:controlled_model}
            \mbf \px = (\mbf{A+BK})\mbf x+\mbf B(\mbf v-\mbf{K\bx}).
          \end{equation}
          The following assumptions will be needed for designing stabilizing controllers $\subss \CC i$.
          \begin{assum}
            \label{ass:stability}
            \begin{enumerate}[(i)]
            \item\label{enu:fischur}The matrices $F_i=A_{ii}+B_iK_i$, $i\in\MM$ are Schur.
            \item\label{enu:fschur}The matrix $\mbf F=\mbf A+\mbf B\mbf K$ is Schur.
            \end{enumerate}
            \begin{flushright}$\blacksquare$\end{flushright}
          \end{assum}

          We discuss now constraints satisfaction. To this purpose, we equip subsystems $\subss\Sigma i$, $i\in\MM$ with the constraints $\subss x i \in \Xset_i,~\subss u i \in \Uset_i$, define the sets $\Xset=\prod_{i\in\MM}\Xset_i$, $\Uset=\prod_{i\in\MM}\Uset_i$ and consider the collective constrained system \eqref{eq:system} with 
          \begin{equation}
            \label{eq:constraints}
            \mbf x \in \Xset, ~
            \mbf u \in \Uset.
          \end{equation}
          As in tube-based MPC control \cite{Mayne2005}, our goal is to compute tightened state constraints $\hXset_i\subseteq\Xset_i$ and input constraints $\Vset_i\subseteq\Uset_i$ guaranteeing that
          \begin{align}
            \label{eq:cnstr_satisfied}
            &\subss \bx i (k)\in\hXset_i,~\subss v i (k)\in\Vset_i,~\forall i\in\MM\\
            &\quad \Rightarrow \mbf{x}(k+1)\in \Xset,~\mbf{u}(k)\in\Uset,\nonumber
          \end{align}
          The next Assumption characterizes the shape of constraints $\Xset_i$, $\hXset_i$, $\Uset_i$ and $\Vset_i$, $i\in\MM$.
          \begin{assum}
            \label{ass:shapesets}
            Constraints $\Xset_i$ and $\hXset_i$ are zonotopes given by
            \begin{equation}
              \label{eq:statezonotope}
\begin{aligned}
              \Xset_i &=\{\subss x i \in\Rset^{n_i}|f_{i,r}^T\subss{x}{i}\leq 1, r\in 1:\bar{r}_i\} = \{ \subss x i \in\Rset^{n_i}|\FF_i\subss{x}{i}\leq \One_{\bar r_i} \\ &= \{\subss x i \in\Rset^{n_i}|\subss x i=\Xi_i d_i,\mbox{ }\norme{d_i}{\infty}\leq 1 \},
\end{aligned}
            \end{equation}
            \begin{equation}
              \label{eq:tightenedstatezonotope}
\begin{aligned}
              \hXset_i &=\{\subss{\hx}{i}\in\Rset^{n_i}|\hf_{i,r}^T\subss{\hx}{i}\leq \hl_i, r\in 1:\bar{\hr}_i\} = \{\subss{\hx}{i}\in\Rset^{n_i}| \hFF_i\subss{\hx}{i}\leq \hl_i \One_{\bar r_i}\} \\ &= \{\subss\hx i=\hXi_i \hd_i,\mbox{ }\norme{\hd_i}{\infty}\leq \hl_i \},
\end{aligned}
            \end{equation}
            where $\FF_i=(f_{i,1}^T,\ldots,f_{i,\bar r_i}^T)\in\Rset^{\bar{r}_i\times n_i}$, $\rank(\FF_i)=n_i$, $d_i\in\Rset^{e_i}$, $\Xi_i\in\Rset^{n_i\times e_i}$, 
$\hl_i\in\Rset_+$, $\hFF_i=(\hf_{i,1}^T,\ldots,\hf_{i,\bar \hr_i}^T)\in\Rset^{\bar{\hr}_i\times n_i}$, $\hd_i\in\Rset^{\bar{e}_i}$ and $\hXi_i\in\Rset^{n_i\times \bar{e}_i}$.\\
            Constraints $\Uset_i$ and $\Vset_i$, $i\in\MM$ are polytopes containing the origin in their interior, that, without loss of generality, are defined as follows
            \begin{equation}   
              \label{eq:inputconstraints}
              \Uset_i = \{\subss{u}{i}\in\Rset^{m_i}|h_{i,r}^T\subss{u}{i}\leq 1, r\in 1:r_{u_i}\} = \{\subss{u}{i}\in\Rset^{m_i}| \HH_i\subss{u}{i}\leq \One_{r_{u_i}} \},
            \end{equation}
            \begin{equation}   
              \label{eq:tightenedinputconstraints}
              \Vset_i = \{\subss{v}{i}\in\Rset^{m_i}|h_{i,r}^T\subss{v}{i}\leq 1-l_{v_{i,r}}, r\in 1:r_{u_i}\} = \{ \subss{v}{i}\in\Rset^{m_i}|\HH_i\subss{v}{i}\leq \One_{r_{u_i}}-l_{v_i} \},
            \end{equation}
            where $\HH_i=(h_{i,1}^T,\ldots,h_{i,r_{u_i}}^T)\in\Rset^{r_{u_i}\times m_i}$, $l_{v_{i,r}}\in\Rset_+$ and $l_{v_i}=(l_{v_{i,1}},\ldots,l_{v_{i,{r_{u_i}}}})$.
            \begin{flushright}$\blacksquare$\end{flushright}
          \end{assum}          
          
          From the results in \cite{Kolmanovsky1998}, under Assumptions \ref{ass:stability}-(\ref{enu:fischur}) and \ref{ass:shapesets} there exist nonempty RPIs $\Zset_i\subseteq\Rset^{n_i}$, $i\in\MM$ for the dynamics
          \begin{equation}
            \label{eq:errorDyn}
            \subss \pz i = (A_{ii}+B_iK_i)\subss z i+\subss w i
          \end{equation}
          and $w_i\in\Wset_i=\bigoplus_{j\in\NN_i}A_{ij}\Xset_j$. In particular, for $\delta_i>0$, we denote with $\Zset_i(\delta_i)$ an RPI set that is a $\delta_i$-outer approximation of the minimal RPI for \eqref{eq:errorDyn} and $\subss w i\in\Wset_i$.\\
          For guaranteeing \eqref{eq:cnstr_satisfied}, we introduce the following Assumption.
          \begin{assum}
            \label{ass:constr_satisf}
            There exist $\delta_i>0$ and nonempty constraint sets $\hXset_i$ and $\Vset_i$, $\forall i\in\MM$ verifying
              \begin{equation}
                \label{eq:stateinclu}
                \hXset_i\oplus\Zset_i(\delta_i)\subseteq\Xset_i
              \end{equation}
              \begin{equation}
                \label{eq:inputinclu}
                \Vset_i\oplus K_i\Zset_i(\delta_i)\subseteq\Uset_i.
              \end{equation}
            \begin{flushright}$\blacksquare$\end{flushright}
          \end{assum}
          Note that, by construction, one has $\Zset_i(\delta_i)\supseteq \Wset_i$ and therefore \eqref{eq:stateinclu} and \eqref{eq:inputinclu} cannot be verified if $\Wset_i$ is ``too big'', i.e. $\Wset_i\supseteq\Xset_i$ or $K_i\Wset_i\supseteq\Uset_i$.

          Under Assumptions \ref{ass:stability}-\ref{ass:constr_satisf}, as in \cite{Mayne2005}, we set in  \eqref{eq:tubecontrol}
          \begin{equation}
            \label{eq:def_kappa_eta}
            \kappa_i(\subss x i(t))=\subss v i(0|t),\qquad\eta_i(\subss x i(t))=\subss\hx i(0|t)
          \end{equation}
          where $\subss v i(0|t)$ and $\subss\hx i(0|t)$ are optimal values of variables $\subss v i(0)$ and $\subss\hx i(0)$, respectively, appearing in the following MPC-$i$ problem to be solved at time $t$
          \begin{subequations}   
            \label{eq:decMPCProblem}
            \begin{align}              
              &\label{eq:costMPCProblem}\Pset_i^N(\subss x i(t)) = \min_{\substack{\subss\hx i(0)\\\subss v i(0:N_i-1)}}\sum_{k=0}^{N_i-1}\ell_i(\subss\hx i(k),\subss v i(k))+V_{f_i}(\subss\hx i(N_i))\\
              &\label{eq:inZproblem}\subss x i(t)-\subss \hx i(0)\in\Zset_i(\delta_i) \\
              &\label{eq:dynproblem}\subss \hx i(k+1)=A_{ii}\subss \hx i(k)+B_i\subss v i(k)\qquad\qquad k\in0:N_i-1 \\              
              &\label{eq:inhXproblem}\subss \hx i(k)\in\hXset_i\qquad\qquad\qquad\qquad\qquad\qquad\qquad k\in0:N_i-1 \\
              &\label{eq:inVproblem}\subss v i(k)\in\Vset_i\qquad\qquad\qquad\qquad\qquad\qquad\qquad k\in0:N_i-1 \\
              &\label{eq:inTerminalSet}\subss \hx i(N_i)\in\hXset_{f_i}
            \end{align}
          \end{subequations}
          In \eqref{eq:decMPCProblem}, $N_i\in\Nset$ is the prediction horizon, $\ell_i(\subss\hx i(k),\subss v i(k)):\Rset^{n_i\times m_i}\rightarrow\Rset_+$ is the stage cost and $V_{f_i}(\subss\hx i(N_i)):\Rset^{n_i}\rightarrow\Rset_+$ is the final cost, fulfilling the following assumption.            
          \begin{assum}
            \label{ass:axiomsMPC}
            For all $i\in\MM$, there exist an auxiliary control law $\kappa_i^{aux}(\subss\hx i)$ and a $\KK_\infty$ function $\BB_i$ such that:
            \begin{enumerate}[(i)]
            \item\label{enu:boundstagecost}$\ell_i(\subss x i,\subss u i)\geq \BB_i(\norme{(\subss x i,\subss u i)}{})$, for all $\subss x i\in\Rset^{n_i}$,  $\subss u i\in\Rset^{m_i}$ and $\ell_i(0,0)=0$;
            \item\label{enu:invariantAux}$\hXset_{f_i}\subseteq\hXset_i$ is an invariant set for $\subss\hpx i=A_{ii}\subss\hx i+B_i\kappa_i^{aux}(\subss\hx i)$;
            \item\label{enu:terminalSetAux}$\forall \subss \hx i\in\hXset_{f_i}$, $\kappa_i^{aux}(\subss\hx i)\in\Vset_i$;
            \item\label{enu:decterminal}$\forall \subss \hx i\in\hXset_{f_i}$, $V_{f_i}(\subss\hpx i)-V_{f_i}(\subss\hx i)\leq-\ell_i(\subss\hx i,\kappa_i^{aux}(\subss\hx i))$.
            \end{enumerate}
            \begin{flushright}$\blacksquare$\end{flushright}
          \end{assum}
          We highlight that there are several methods, discussed e.g. in \cite{Rawlings2009}, for computing $\ell_i(\cdot)$, $V_{f_i}(\cdot)$ and $\Xset_{f_i}$ verifying Assumption \ref{ass:axiomsMPC}.

          The next Theorem, that is proved in Appendix \ref{sec:proofstabcon}, provides the main results on stability of the closed-loop system \eqref{eq:controlled_model} and \eqref{eq:def_kappa_eta} equipped with constraints \eqref{eq:constraints}.
          \begin{thm}
            \label{thm:stabcon}
            Let Assumptions \ref{ass:stability}-\ref{ass:axiomsMPC} hold. Define the feasibility region for the MPC-$i$ problem as
            $$
              \Xset_i^N=\{\subss s i\in\Xset_i:~\mbox{~\eqref{eq:decMPCProblem} is feasible for}~\subss x i(t)=\subss s i\}
            $$
            and the collective feasibility region as $\Xset^N=\prod_{i\in\MM}\Xset_i^N$.\\
            Then
            \begin{enumerate}[(i)]
            \item if $\mbf x(0)\in\Xset^N$, i.e. $\subss x i(0)\in\Xset_i^N$ for all $i\in\MM$, constraints \eqref{eq:constraints} are fulfilled at all time instants;
            \item the origin of the closed-loop system \eqref{eq:controlled_model} and \eqref{eq:def_kappa_eta} is asymptotically stable and $\Xset^N$ is a region of attraction.
            \end{enumerate}
            \begin{flushright}$\blacksquare$\end{flushright}
          \end{thm}

          In order to design a DeMPC scheme based on MPC-$i$ problems \eqref{eq:decMPCProblem} and for which Theorem \ref{thm:stabcon} applies, the main problem that still has to be solved is the following one.
          \subsubsection*{Problem $\PP$}
               Compute matrices $K_i$, $i\in\MM_i$, if they exist, verifying Assumptions \ref{ass:stability} and \ref{ass:constr_satisf}. 
               \begin{flushright}$\blacksquare$\end{flushright}

          In the next Section, we show how to solve Problem $\PP$ in a distributed fashion under Assumption \ref{ass:shapesets} complemented by the next assumption.
          \begin{assum}
            \label{ass:givenpara}
            Matrices $\hFF_i$ (and hence $\hXi_i$) in \eqref{eq:tightenedstatezonotope} are given for $i\in\MM$.
            \begin{flushright}$\blacksquare$\end{flushright}
          \end{assum}
          Note that Assumption \ref{ass:givenpara} fixes the shape of set $\hXset_i$, $i\in\MM$ leaving the freedom to choose the zooming parameters $\hl_i$. Also the shape of each set $\Vset_i$ is fixed and, from Assumption \ref{ass:shapesets}, it coincides with the shape of $\Uset_i$ while parameters $l_{v_i}$ are free.

          \section{Decentralized synthesis of DeMPC}
          \label{sec:main}
          The next Theorem will allow us to solve Problem $\PP$ in a distributed fashion.
          \begin{thm}
            \label{thm:main}
            Let Assumptions \ref{ass:shapesets} and \ref{ass:givenpara} hold. For given matrices $K_i$, $i\in\MM$, verifying Assumption \ref{ass:stability}-(\ref{enu:fischur}), if the following conditions are fulfilled
            \begin{equation}
              \label{eq:pseudoinequalities}
              \alpha_i =\sum_{j\in\NN_i}\sum_{k=0}^{\infty}\norme{\FF_iF_i^kA_{ij}\FF_j^\flat}{\infty}<1,~\forall i\in\MM
            \end{equation}
            then
            \begin{enumerate}[(i)]
            \item Assumption \ref{ass:stability}-(\ref{enu:fschur}) holds.
            \item For all $i\in\MM$, defining
              \begin{equation}
                \label{eq:stateinequalitieshatL}
                \underline{\hat L}_{i,r}=\frac{1-\sum_{j\in\NN_i}\sum_{k=0}^{\infty}\norme{f_{i,r}^TF_i^kA_{ij}\Xi_j}{\infty}}{\norme{f_{i,r}^T\hXi_i}{\infty}},\qquad r\in1:\bar{r}_i
              \end{equation}
              there is $\delta_i>0$ such that
              \begin{equation}
                \label{eq:stateinequalities}
                {\hat L}_i = \min_{\substack{r\in1:\bar{r}_i}}~\underline{\hat L}_{i,r}-\frac{\norme{f_{i,r}^T}{\infty}\delta_i}{\norme{f_{i,r}^T\hXi_i}{\infty}}>0.
              \end{equation}
              Furthermore, choosing $\hl_i\in(0,\hat L_i]$ and the set $\hXset_i$ as in \eqref{eq:tightenedstatezonotope}, the inclusion \eqref{eq:stateinclu} holds.
            \item For $\delta_i>0$ verifying \eqref{eq:stateinequalities} assume the following condition is fulfilled
              \begin{equation}
                \label{eq:inputinequalities}
                \beta_i(\delta_i) = \max_{\substack{~r\in1:r_{u_i}}}~\hl_{v_{i,r}}(\delta_i)< 1
              \end{equation}
              with 
              \begin{equation}
                \label{eq:inputless}
                \hl_{v_{i,r}}(\delta_i)=\sup_{\substack{z_i\in\Zset_i(\delta_i)}}~h_{i,r}^TK_iz_i,~r\in1:r_{u_i}.
              \end{equation}
             Then, choosing $\Vset_i$ as in \eqref{eq:tightenedinputconstraints} for $l_{v_{i,r}}=\hl_{v_{i,r}}(\delta_i)$ the inclusion \eqref{eq:inputinclu} holds.
            \end{enumerate}
            \begin{flushright}$\blacksquare$\end{flushright}
          \end{thm}           
          The proof of Theorem \ref{thm:main} can be found in Appendix \ref{sec:proofmain}.
          
          We highlight that under Assumption \ref{ass:givenpara}, for a given $i\in\MM$, the quantities $\alpha_i$ in \eqref{eq:pseudoinequalities}, $\hat L_i$ in \eqref{eq:stateinequalities} and $\beta_i(\delta_i)$ in \eqref{eq:inputinequalities} depend only upon local fixed parameters $\{A_{ii},B_{i},\FF_i,\HH_i\}$, neighbors' fixed parameters $\{A_{ij},\Xi_j\}_{j\in\NN_i}$ (or equivalently $\{A_{ij},\FF_j\}_{j\in\NN_i}$) and local tunable parameters $\{K_i,\delta_i\}$ but not on neighbors' tunable parameters. Moreover, also the computation of sets $\Zset_i(\delta_i)$ depends upon the same parameters. This implies that the choice of $\{K_i,\delta_i\}$ does not influence the choice of $\{K_j,\delta_j\}_{j\neq i}$ and therefore Problem $\PP$ is decomposed in the following independent problems for $i\in\MM$.
          \subsubsection*{Problem $\PP_i$}
               Check if there exist $K_i$ and $\delta_i>0$ such that $\alpha_i<1$, $\hat L_i>0$ an $\beta_i(\delta_i)< 1$.
               \begin{flushright}$\blacksquare$\end{flushright}

          According to Theorem \ref{thm:main}, the solution to Problem $\PP_i$ enables the computation of sets $\hXset_i$ and $\Vset_i$ and therefore the decentralized design of controller MPC-$i$. The overall procedure for the decentralized synthesis of local controllers $\subss \CC i,~i\in\MM$ is summarized in Algorithm \ref{alg:distrisynt}, whose computational aspects are discussed in Section \ref{sec:computational}.

          \begin{algorithm}
            \caption{Design of controller $\subss\CC i$ for system $\subss \Sigma i$}
            \label{alg:distrisynt}
            \textbf{Input}: $A_{ii}$, $B_i$, $\Xset_i$, $\Uset_i$, $\NN_i$, $\{A_{ij}\}_{j\in\NN_i}$, $\{\Xset_{j}\}_{j\in\NN_i}$\\
            \textbf{Output}: controller $\subss \CC i$ in \eqref{eq:tubecontrol}\\
            \begin{enumerate}[1)]
            \item\label{enu:feasProb} Find $K_i$ and $\delta_i>0$ such that Assumption \ref{ass:stability}-(\ref{enu:fischur}) is fulfilled, $\alpha_i<1$, \eqref{eq:stateinequalities} holds and $\beta_i(\delta_i)<1$. If they do not exist \textbf{stop} (the controller $\subss\CC i$ cannot be designed).
            \item Compute sets $\Wset_i=\bigoplus_{j\in\NN_i}A_{ij}\Xset_j$ and $\Zset_i(\delta_i)$.
            \item Compute $\hat L_i$ as in \eqref{eq:stateinequalities}, choose $\hl_i=\hat L_i$ and define $\hXset_i$ as in \eqref{eq:tightenedstatezonotope}.
            \item Compute $\hl_{v_{i,r}}(\delta_i)$ as in \eqref{eq:inputless}, set $l_{v_{i,r}}=\hl_{v_{i,r}}(\delta_i)$ and define $\Vset_i$ as in \eqref{eq:tightenedinputconstraints}.
            \item Compute $\ell_i(\cdot)$, $V_{f_i}(\cdot)$ and $\Xset_{f_i}$ verifying Assumption \ref{ass:axiomsMPC}. 
            \end{enumerate}
          \end{algorithm}

          In view of the previous discussion, the link between controllers designed through Algorithm \ref{alg:distrisynt} and Theorem \ref{thm:stabcon} can be summarized as follows.
          \begin{prop}
            \label{prop:asshold}
            Under Assumptions \ref{ass:shapesets} and \ref{ass:givenpara}, if for all $i\in\MM$ controllers $\subss\CC i$ are designed according to Algorithm \ref{alg:distrisynt}, then all Assumptions of Theorem \ref{thm:stabcon} are fulfilled.
          \end{prop}

     \section{Plug-and-play operations}
          \label{sec:plugplay}
          In this Section we discuss the synthesis of new controllers and the redesign of existing ones when subsystems are added to or removed from system \eqref{eq:subsystem}. The goal will be to preserve stability of the origin and constraint satisfaction for the new closed-loop system. Note that plugging in and unplugging of subsystems are here considered as off-line operations. Therefore, the overall plant is not modeled as a switching system. As a starting point, we consider a plant composed by subsystems $\subss \Sigma i$, $i\in\MM$ equipped with local controllers $\subss \CC i$, $i\in\MM$ produced by Algorithm \ref{alg:distrisynt}.

          \subsection{Plugging in operation}
               \label{sec:plugin}
               We start considering the plugging of subsystem $\subss{\Sigma}{M+1}$, characterized by parameters $A_{M+1~M+1}$, $B_{M+1}$, $\Xset_{M+1}$, $\Uset_{M+1}$, $\NN_{M+1}$ and $\{A_{ij}\}_{j\in\NN_{M+1}}$, into an existing plant. In particular $\NN_{M+1}$ identifies the subsystems that will be physically coupled to $\subss{\Sigma}{M+1}$ and $\{A_{ij}\}_{j\in\NN_{M+1}}$ are the corresponding coupling terms. For building the controller $\subss{\CC}{M+1}$ we execute Algorithm \ref{alg:distrisynt} that needs information only from systems $\subss{\Sigma}{j}$, $j\in\NN_{M+1}$. If Algorithm \ref{alg:distrisynt} stops before the last step we declare that $\subss{\Sigma}{M+1}$ cannot be plugged in. Let $\SSS_i=\{j:i\in\NN_j\}$ be the set of successors to system $i$. Since each system $\subss{\Sigma}{j}$, $j\in\SSS_{M+1}$ has the new neighbor $\subss{\Sigma}{M+1}$, it can be happen that existing matrices $K_j$, $j\in\SSS_{M+1}$ now give $\alpha_j\geq 1$ or $\hat L_j\leq 0$ or $\beta_i(\delta_i)\geq 1$. Indeed, when $\NN_j$ gets larger, the quantity $\alpha_j$ in \eqref{eq:pseudoinequalities} (respectively $\hat L_j$ in \eqref{eq:stateinequalities}) can only increase (respectively decrease). Furthermore, the size of the set $\Zset_j(\delta_j)$ increases and therefore the condition in \eqref{eq:inputinequalities} could be violated. This means that for each $j\in\SSS_{M+1}$ the controllers $\subss\CC j$ must be redesigned according to Algorithm \ref{alg:distrisynt}. Again, if Algorithm \ref{alg:distrisynt} stops before completion for some $j\in\SSS_{M+1}$, we declare that $\subss{\Sigma}{M+1}$ cannot be plugged in.\\
               In conclusion, the addition of system $\subss{\Sigma}{M+1}$ triggers the design of controller $\subss{\CC}{M+1}$ and the redesign of controllers $\subss\CC j$, $j\in\SSS_{M+1}$ according to Algorithm \ref{alg:distrisynt}. Note that controller redesign does not propagate further in the network, i.e. even without changing controllers $\subss\CC i$, $i\notin\{M+1\}\bigcup\SSS_{M+1}$ stability of the origin and constraint satisfaction are guaranteed for the new closed-loop system.
          
          \subsection{Unplugging operation}
               \label{sec:unplug}
               We consider the unplugging of  system $\subss{\Sigma}{k}$, $k\in\MM$. Since for each $i\in\SSS_k$ the set $\NN_i$ gets smaller, we have that $\alpha_i$ in \eqref{eq:pseudoinequalities} (respectively $\hat L_i$ in \eqref{eq:stateinequalities}) cannot increase (respectively cannot decrease). Furthermore, the size of the set $\Zset_i(\delta_i)$ cannot increase and therefore the inequality \eqref{eq:inputinequalities} cannot be violated. This means that for each $i\in\SSS_k$ the controller $\subss\CC i$ does not have to be redesigned. Moreover since for each system $\subss\Sigma j$, $j\notin \{k\}\bigcup\SSS_k$ the set $\NN_j$ does not change, the redesign of controller $\subss\CC j$ is not required.\\
               In conclusion, removal of system $\subss\Sigma k$ does not require the redesign of any controller, in order to guarantee stability of the origin and constraints satisfaction for the new closed-loop system. However we highlight that since systems $\subss\Sigma i$, $i\in\SSS_k$ have one neighbor less, the redesign of controllers $\subss\CC i$ through Algorithm \ref{alg:distrisynt} could improve the performance.

     \section{Practical design and computational aspects}
          \label{sec:computational}

          \subsection{Automatic design of $K_i$ and $\delta_i$}
               \label{sec:autoKi}
               The most difficult part of Algorithm \ref{alg:distrisynt} is step \ref{enu:feasProb} and in this Section we propose an automatic method for computing the matrix $K_i$ and $\delta_i>0$. We assume that $K_i$ is the LQ controller associated to matrices $Q_i\geq0$ and $R_i>0$, i.e.
               \begin{equation}
                 \label{eq:kilq}
                 K_i=(R_i+B_i^T\bar{P}_iB_i)^{-1}B_i^T\bar{P}_iA_{ii}
               \end{equation}
               where $\bar{P}_i$ is the solution of the stationary Riccati equation
               $$
               A_{ii}^T\bar{P}_iA_{ii}+Q_i-A_{ii}^T\bar{P}_iB_i(R_i+B_i^T\bar{P}_iB_i)^{-1}B_i^T\bar{P_i}A_{ii}=\bar{P}_i.
               $$
               We then solve the following nonlinear optimization problem
               \begin{subequations}   
                 \label{eq:optimKi}
                 \begin{align}                  
                   &\label{eq:costoptimKi}\min_{\substack{\delta_i,~Q_i,~R_i}}~\mu_{\alpha_i}\alpha_i+\mu_{\beta_i}\beta_i(\delta_i)\\
                   &\label{eq:QRoptimKi}Q_i\geq 0,~R_i>0\\
                   &\delta_i>0\\
                   &\label{eq:alphaLoptimKi}\alpha_i<1,~\hat L_i>0\\
                   &\label{eq:betaoptimKi}\beta_i(\delta_i)<1
                 \end{align}
               \end{subequations}
               where $\mu_{\alpha_i}\geq 0$ and $\mu_{\beta_i}\geq 0$.

               Feasibility of problem \eqref{eq:optimKi} guarantees that Algorithm \ref{alg:distrisynt} does not stop and then the controller $\subss\CC i$ can be successfully designed. Moreover, in \eqref{eq:costoptimKi} weights $\mu_{\alpha_i}$ and $\mu_{\beta_i}$ establish a trade-off between the maximization of sets $\hXset_i$ and $\Vset_i$, respectively. A few remarks on the computations required for solving \eqref{eq:optimKi} are in order. First, beside the computation of  $K_i$ as in \eqref{eq:kilq}, problem \eqref{eq:optimKi} requires the computation of the set $\Zset_i(\delta_i)$ that can be done using methods in \cite{Rakovic2005a}, simplified as follows. Under Assumption \ref{ass:shapesets}, $\Wset_i=\oplus_{j\in\NN_i}A_{ij}\Xset_j$ is a zonotope set defined as $\Wset_i=\{\subss w i=\Xi_{w_i}d_{w_i},~\norme{d_{w_i}}{\infty}\leq 1\}$. Hence, using the procedure proposed in \cite{Rakovic2005a}, the set $\Zset_i(\delta_i)$ is also a zonotope, defined as $\Zset_i(\delta_i)=\{\subss z i=\Xi_{z_i}d_{z_i},~\norme{d_{z_i}}{\infty}\leq 1\}$, where $\Xi_{z_i}=\matr{\Xi_{w_i} & F_{ii}\Xi_{w_i} & \ldots & F_{ii}^{s_i-1}\Xi_{w_i}}$ with $s_i$ computed using Algorithm \ref{alg:distrisynt} in \cite{Rakovic2005a}. Since $\Wset_i$ and $\Zset_i(\delta_i)$ are zonotopes, using \eqref{eq:suppfunc}, we can explicitly calculate the support function used in Algorithm \ref{alg:distrisynt} in \cite{Rakovic2005a} and rewrite \eqref{eq:inputless} as
               $$
               \hl_{v_{i,r}}(\delta_i)=\norme{\Xi_{z_i}^TK_i^Th_{i,r}}{1},~\forall r\in 1:r_{u_i}.
               $$
               Second, we highlight that in absence of input constraints $\Uset_i$, constraint \eqref{eq:betaoptimKi} (and hence the computation of RPI sets $\Zset_i(\delta_i)$) is not necessary. Indeed if $\Uset_i=\Rset^{m_i}$, the inclusion \eqref{eq:inputinclu} holds for all sets $\Vset_i\subseteq\Rset^{m_i}$. Third, the series in \eqref{eq:pseudoinequalities} and \eqref{eq:stateinequalitieshatL} involve only positive terms and can be easily truncated either if \eqref{eq:alphaLoptimKi} is violated or if summands fall below the machine precision. Finally, in order to simplify problem \eqref{eq:optimKi} one can assume $Q_i=\diag(q_{i,1},\ldots,q_{i,n_i})$ and $R_i=\diag(r_{i,1},\ldots,r_{i,m_i})$ hence replacing the matrix inequalities in \eqref{eq:QRoptimKi} with the scalar inequalities $q_{i,k}\geq 0$, $k\in 1:n_i$ and $r_{i,k}> 0$, $k\in 1:m_i$.

          \subsection{Parameter-dependent subsystem}
               \label{sec:parametersdep}
               In many engineering applications parameters of subsystem $i$ are influenced by neighboring subsystems. We model this scenario replacing \eqref{eq:subsystem} with
               \begin{equation}
                 \label{eq:subsystemparameters}
                 \subss\Sigma i^p:\quad\subss \px i=A_{ii}(\xi_{ii},\{\xi_{ij}\}_{j\in\NN_i})\subss x i+B_i(\xi_{ii},\{\xi_{ij}\}_{j\in\NN_i})\subss u i+\sum_{j\in\NN_i}A_{ij}\subss x j
               \end{equation}
               where $\xi_{ij}\in\Rset^{p_{ij}}$ are parameter vectors.\\
               We highlight that for a given sets $\NN_i$, $i\in\MM$, matrices $A_{ii}$ and $B_i$ are constant and design of P\&P DeMPC regulators can be still done using the methods described in Section \ref{sec:main}. Furthermore, the procedure for plugging in a new system discussed in Section \ref{sec:plugin} can be applied with no change since it requires the redesign of controllers $\subss\CC j$, $j\in\SSS_{M+1}$, i.e. controllers associated to the subsystems $\subss\Sigma j^p$ for which matrices $A_{jj}$ and $B_j$ could change. However, when system $\subss\Sigma k^p$ gets unplugged, it is now mandatory to retune all controllers $\subss\CC j$, $j\in\SSS_k$ since changes of matrices $A_{jj}$ and $B_j$ could hamper the fulfillment of conditions \eqref{eq:pseudoinequalities}, \eqref{eq:stateinequalities} or \eqref{eq:inputinequalities} when using the matrices $K_j$ and the scalars $\delta_j$ computed prior to the subsystem removal. Moreover, if Algorithm \ref{alg:distrisynt} stops before completing the redesign of controllers $\subss\CC j$, $\forall j\in\SSS_k$, we declare that subsystem $\subss\Sigma k^p$ cannot be unplugged.

     \section{Example: Power Network System}
          \label{sec:example}
          In this Section, we apply the proposed DeMPC scheme to a power network system composed by several power generation areas coupled through tie-lines. We aim at designing the AGC layer with the goals of 
          \begin{itemize}
          \item keeping the frequency approximately at a nominal value;
          \item controlling the tie-line powers in order to reduce power exchanges between areas. In the asymptotic regime each area should compensate for local load steps and produce the required power.
          \end{itemize}
          In particular we will show advantages brought about by P\&P DeMPC when generation areas are connected/disconnected to/from an existing network.
          
          The dynamics of an area equipped with primary control and linearized around equilibrium value for all variables can be described by the following continuous-time LTI model \cite{Saadat2002}
          \begin{equation}
            \label{eq:ltipower}
            \subss{\Sigma}{i}^C:\quad\subss{\dot{x}}{i} = A_{ii}\subss x i + B_{i}\subss u i + L_{i}\Delta P_{L_i} + \sum_{j\in\NN_i}A_{ij}\subss x j
          \end{equation}
          where $\subss x i=(\Delta\theta_i,~\Delta\omega_i,~\Delta P_{m_i},~\Delta P_{v_i})$ is the state, $\subss u i = \Delta P_{ref_i}$ is the control input of each area, $\Delta P_{L}$ is the local power load and $\NN_i$ is the sets of neighboring areas, i.e. areas directly connected to $\subss\Sigma i^C$ through tie-lines. The matrices of system \eqref{eq:ltipower} are defined as
          \begin{equation}
            \label{eq:matrixpower}
            \begin{aligned}
              A_{ii}(\{P_{ij}\}_{j\in\NN_i}) &= \matr{ 0 & 1 & 0 & 0 \\ -\frac{\sum_{j\in\NN_i}{P_{ij}} }{2H_i} & -\frac{D_i}{2H_i} & \frac{1}{2H_i} & 0 \\ 0 & 0 & -\frac{1}{T_{t_i}}  & \frac{1}{T_{t_i}} \\ 0 & -\frac{1}{R_iT_{g_i}} & 0 & -\frac{1}{T_{g_i}} }
              &B_{i} = \matr{ 0 \\ 0 \\ 0 \\ \frac{1}{T_{g_i}} }\\
              A_{ij} &= \matr{ 0 & 0 & 0 & 0 \\ \frac{P_{ij}}{2H_i} & 0 & 0 & 0 \\ 0 & 0 & 0  & 0 \\ 0 & 0 & 0 & 0 }
              &L_{i} = \matr{ 0 \\ -\frac{1}{2H_i} \\ 0 \\ 0 }
            \end{aligned}
          \end{equation}
          For the meaning of constants as well as parameter values we defer the reader to Appendix \ref{sec:valexe}. We highlight that all parameter values are within the range of those used in Chapter 12 of \cite{Saadat2002}.\\
          We note that model~\eqref{eq:ltipower} is input decoupled since both $\Delta P_{ref_i}$ and $\Delta P_{L_i}$ act only on subsystem $\subss{\Sigma}{i}^C$. Moreover, subsystems $\subss\Sigma i^C$ are parameter dependent since the local dynamics depends on the quantities $-\frac{\sum_{j\in\NN_i}{P_{{ij}}} }{2H_i}$. We equip each subsystem $\subss{\Sigma}{i}^C$ with the constraints on $\Delta\theta_i$ and on $\Delta P_{ref_i}$ specified in Appendix \ref{sec:valexe}. We obtain models $\subss\Sigma i$ by discretizing models $\subss\Sigma i^C$ with $1~sec$ sampling time, using exact discretization and treating $\subss u i$, $\Delta P_{L_i}$, $\subss x j,~j\in\NN_i$ as exogenous signals.\\
          In the following we first design the AGC layer for a power network composed by four areas (Scenario 1) and then we show how in presence of connection/disconnection of an area (Scenario 2 and 3, respectively) the AGC can be redesigned via plugging in and unplugging operations.
    
          \subsection{Scenario 1}  
               \label{sec:scenario1}
\begin{figure}[!ht]
                 \centering
                 \includegraphics[scale=0.75]{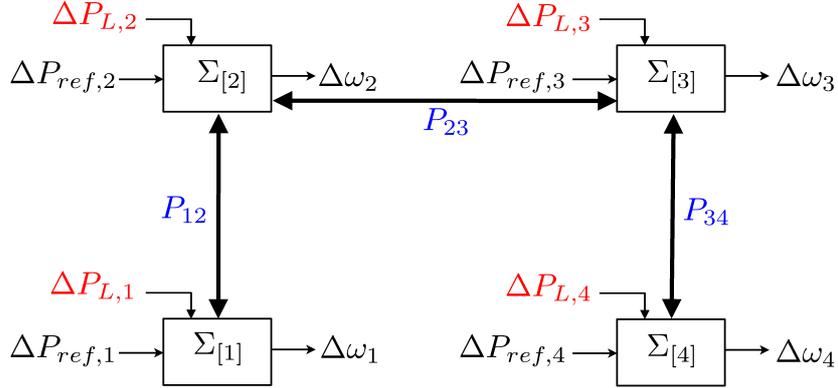}
                 \caption{Power network system of Scenario 1}
                 \label{fig:scenario1}
               \end{figure}

\begin{figure}[!ht]
                 \centering
                 \includegraphics[scale=0.5]{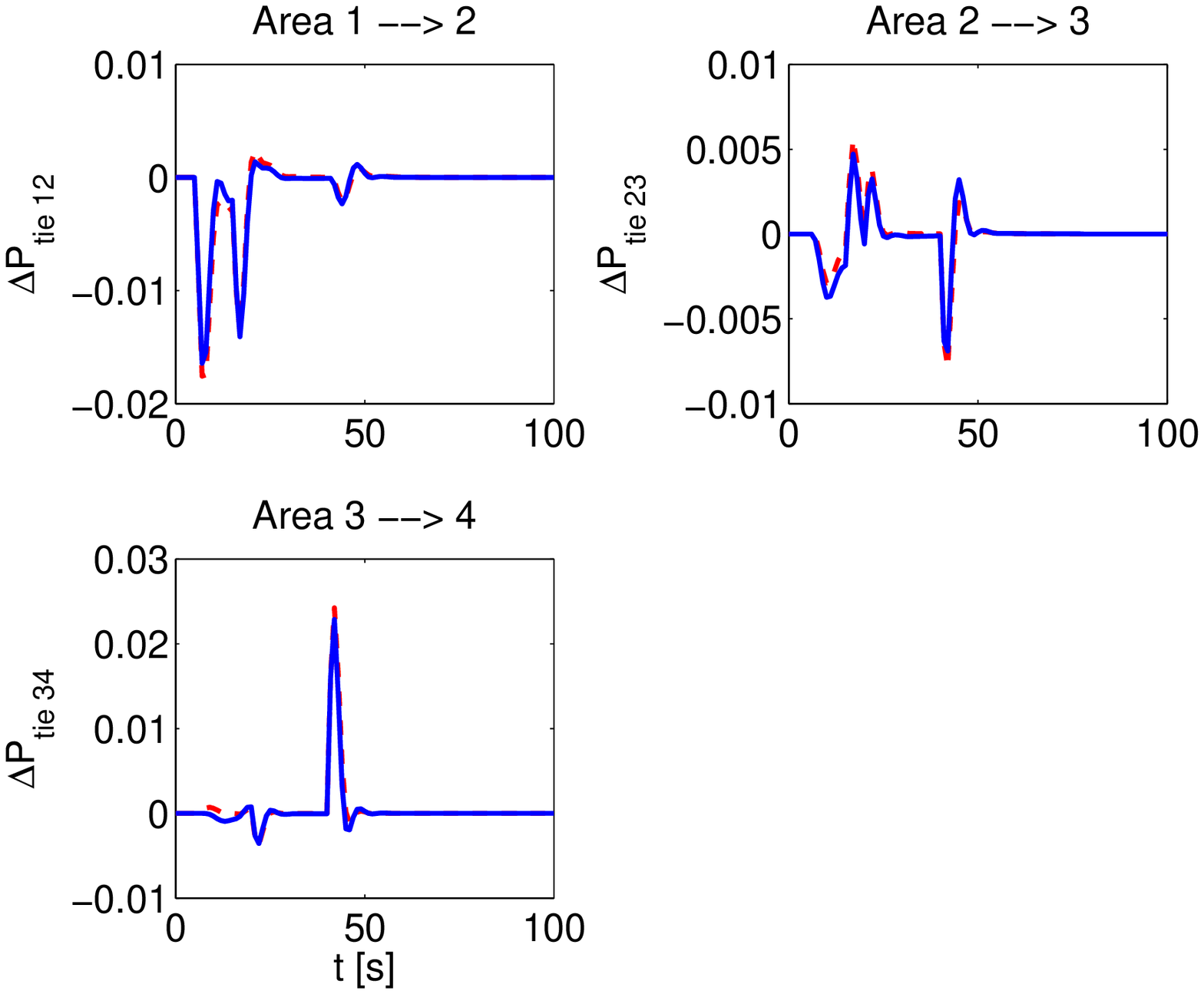}
                 \caption{Simulation Scenario 1: tie-line power between each area controlled by the proposed DeMPC (bold line) and centralized MPC (dashed line).}
                 \label{fig:simulationscen1tiepower}
               \end{figure}

               We consider four areas interconnected as in Figure~\ref{fig:scenario1}.
               For each system $\subss\Sigma i$ we synthesize the controller $K_i,~i\in\MM$ solving an LQ problem for the nominal system, as shown in Section \ref{sec:autoKi} with $\mu_{\alpha_i}=1$ and $\mu_{\beta_i}=1$, $\forall i\in\MM$, and obtain the following matrices
               \begin{equation}
                 \label{eq:controllersK}
                   \oneblock{
                     K_1 &= -\matr{ 0.508 & 0.201 & 0.006 & 0.001 },~K_2 &= -\matr{ 0.729 & 0.437 & 0.008 & 0.002 }, \\
                     K_3 &= -\matr{ 3.409 & 4.759 & 0.090 & 0.030 },~K_4 &= -\matr{ 4.426 & 6.348 & 0.233 & 0.038 }, 
                   }
               \end{equation}
               that allow inequalities \eqref{eq:pseudoinequalities} to be fulfilled. Hence $\mbf K$ verifies Assumption \ref{ass:stability}-(\ref{enu:fschur}). Setting $\delta_i=10^{-4},~\forall i\in\MM$ and applying steps 2-5 of Algorithm \ref{alg:distrisynt}, we can compute sets $\Zset_i(\delta_i)$, $\hXset_i$ and $\Vset_i$ such that inclusions \eqref{eq:stateinclu} and \eqref{eq:inputinclu} hold, $\forall i\in\MM$. Control variables $\subss u i$ are obtained through \eqref{eq:tubecontrol} where $\subss v i=\kappa_i(\subss x i)$ and $\subss\bx i=\eta_i(\subss x i)$ are computed at each time $t$ solving the optimization problem \eqref{eq:decMPCProblem} and replacing the cost function in \eqref{eq:costMPCProblem} with the following one depending upon $\subss x i^O=(0,~0,~\Delta P_{L_i},~\Delta P_{L_i})$ and $\subss u i^O=\Delta P_{L_i}$
               \begin{equation}
                 \label{eq:decMPCproblemPower}
                 J_i^{N_i} = \sum_{k=t}^{t+N_i-1}(\norme{\subss\hx i(k)-\subss x i^O}{\hat Q_i}+\norme{\subss v i(k)-\subss u i^O}{\hat R_i})+\norme{\subss x i(t+N_i)-\subss x i^O}{\hat S_i}.
               \end{equation}
               Note that, except for the above modification of the cost function, that is needed for counteracting load disturbances, we followed exactly the design procedure described in Section \ref{sec:modelcontroller}. Moreover, we highlight that each area can locally absorb the load steps specified in Table \ref{tab:simulationscen1} of Appendix \ref{sec:valexe}. This is also shown by convergence to zero of the power transfer between areas $i$ and $j$ given by
               \begin{equation}
                 \label{eq:powerexchanged}
                 \Delta P_{{tie}_{ij}} = P_{ij}(\Delta\theta_i-\Delta\theta_j)
               \end{equation}
               and represented in Figure \ref{fig:simulationscen1tiepower}.
               \begin{figure}[!ht]
                 \centering
                 \subfigure[\label{fig:simulationscen1freq}Frequency deviation in each area controlled by the proposed De-MPC (bold line) and centralized MPC (dashed line).]{\includegraphics[scale=0.5]{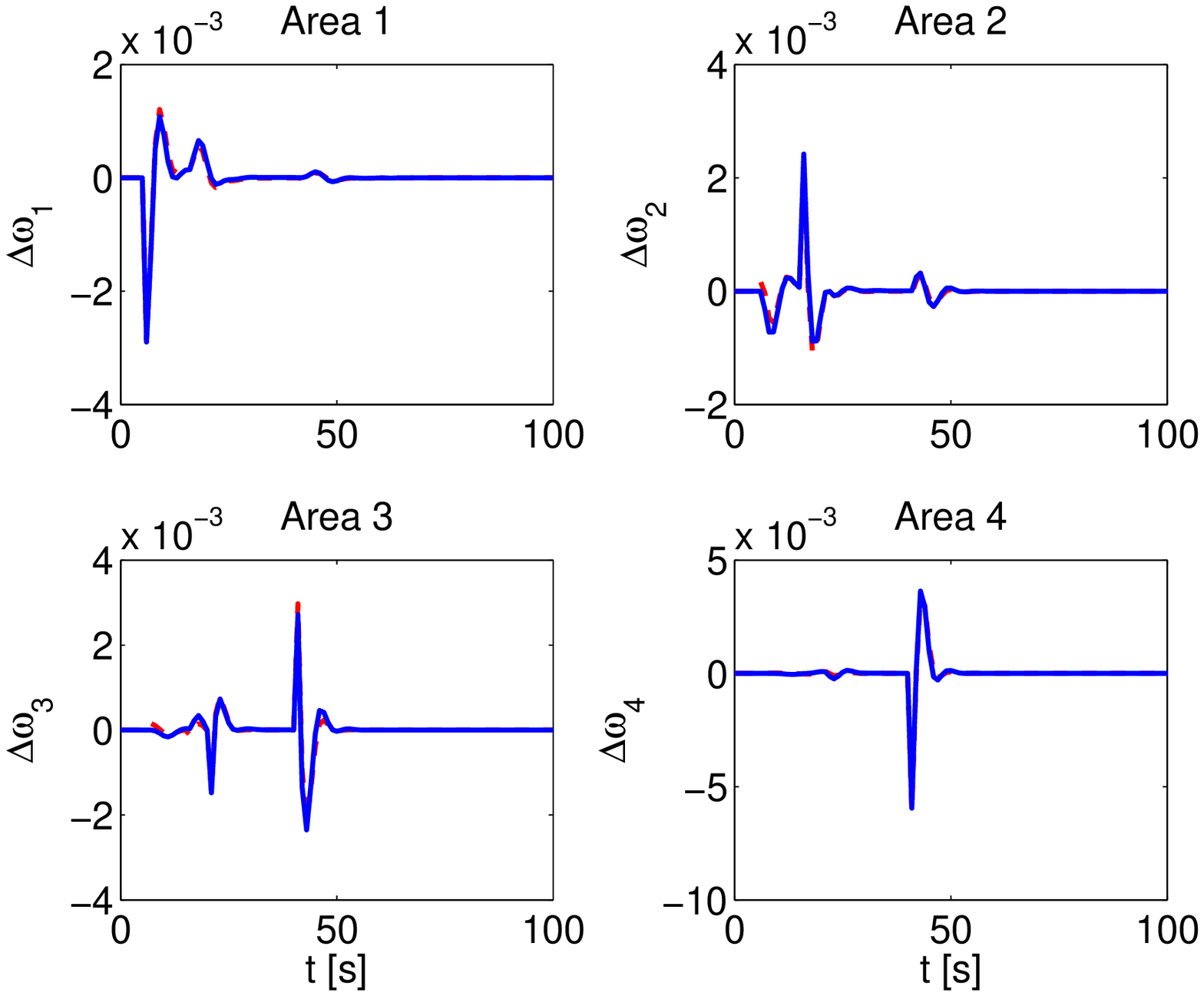}}\\
                 \subfigure[\label{fig:simulationscen1ref}Load reference set-point in each area controlled by the proposed De-MPC (bold line) and centralized MPC (dashed line).]{\includegraphics[scale=0.5]{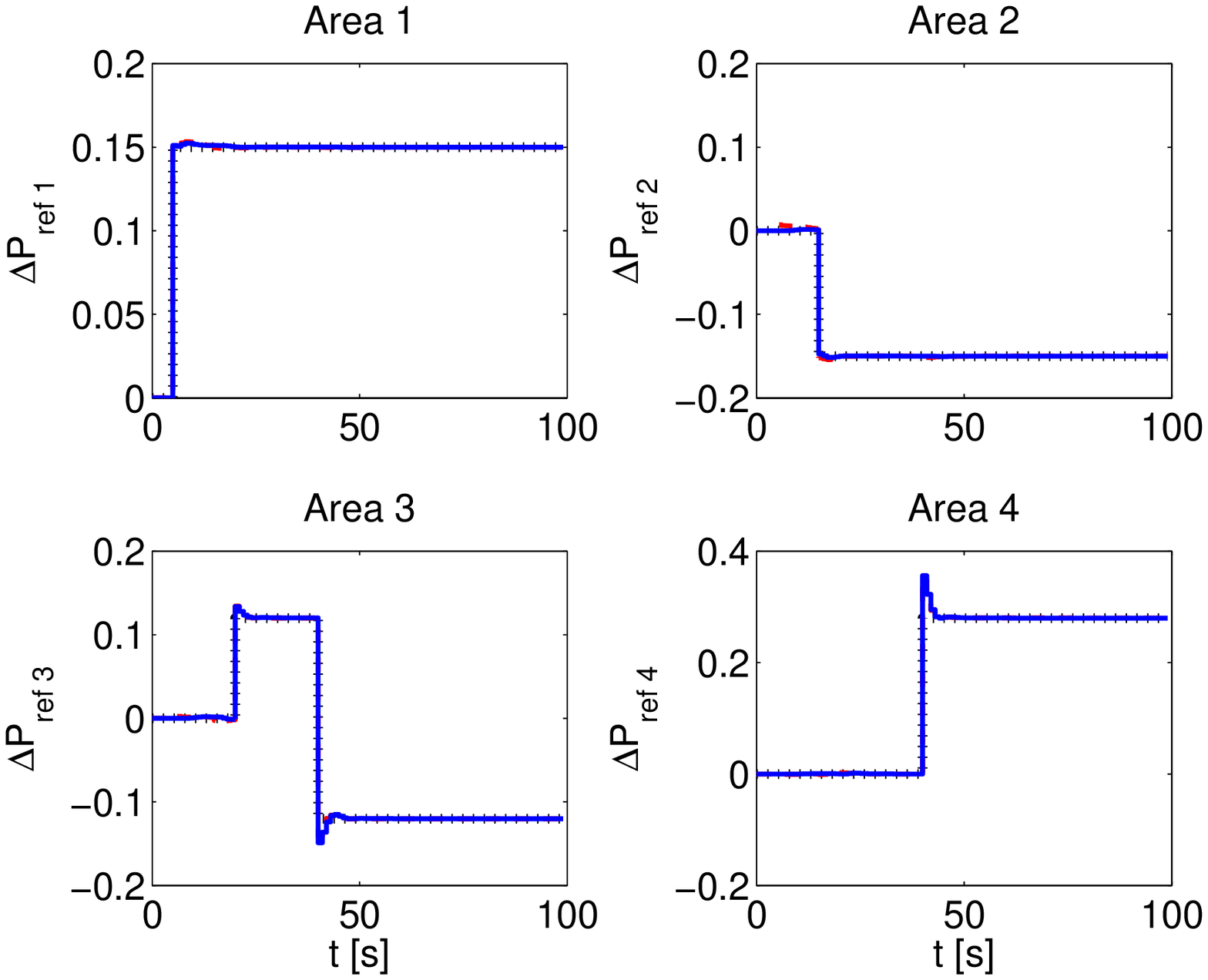}}
                 \caption{Simulation Scenario 1: \ref{fig:simulationscen1freq} Frequency deviation and \ref{fig:simulationscen1ref} Load reference in each area.}
                 \label{fig:simulationscen1}
               \end{figure}

               In Figure~\ref{fig:simulationscen1} we compare the performance of proposed DeMPC with the performance of centralized MPC. For centralized MPC we consider the overall system composed by the four areas, use the cost function $\sum_{i\in\MM}J_i^N$ and impose the collective constraints \eqref{eq:constraints}. The prediction horizon is $N_i=20,~i\in\MM$ for MPC-$i$ controllers and $N=20$ for centralized MPC. In the control experiment, step power loads $\Delta P_{L_i}$ specified in Appendix \ref{sec:valexe} have been used and they account for the step-like changes of the control variables in Figure \ref{fig:simulationscen1}. We highlight that the performance of decentralized and centralized MPC are totally comparable, in terms of frequency deviation (Figure~\ref{fig:simulationscen1freq}), control variables (Figure~\ref{fig:simulationscen1ref}) and power transfers $\Delta P_{{tie}_{ij}}$ (Figure \ref{fig:simulationscen1tiepower}).

          \subsection{Scenario 2}
               \label{sec:scenario2}
               We consider the power network proposed in Scenario 1 and we add a fifth area connected as in Figure \ref{fig:scenario2} with values of parameters and constraints listed in Table \ref{tab:scenario123} of Appendix \ref{sec:valexe}. Therefore, the set of successors to system $5$ is $\SSS_5=\{2,4\}$.
               \begin{figure}[!ht]
                 \centering
                 \includegraphics[scale=0.75]{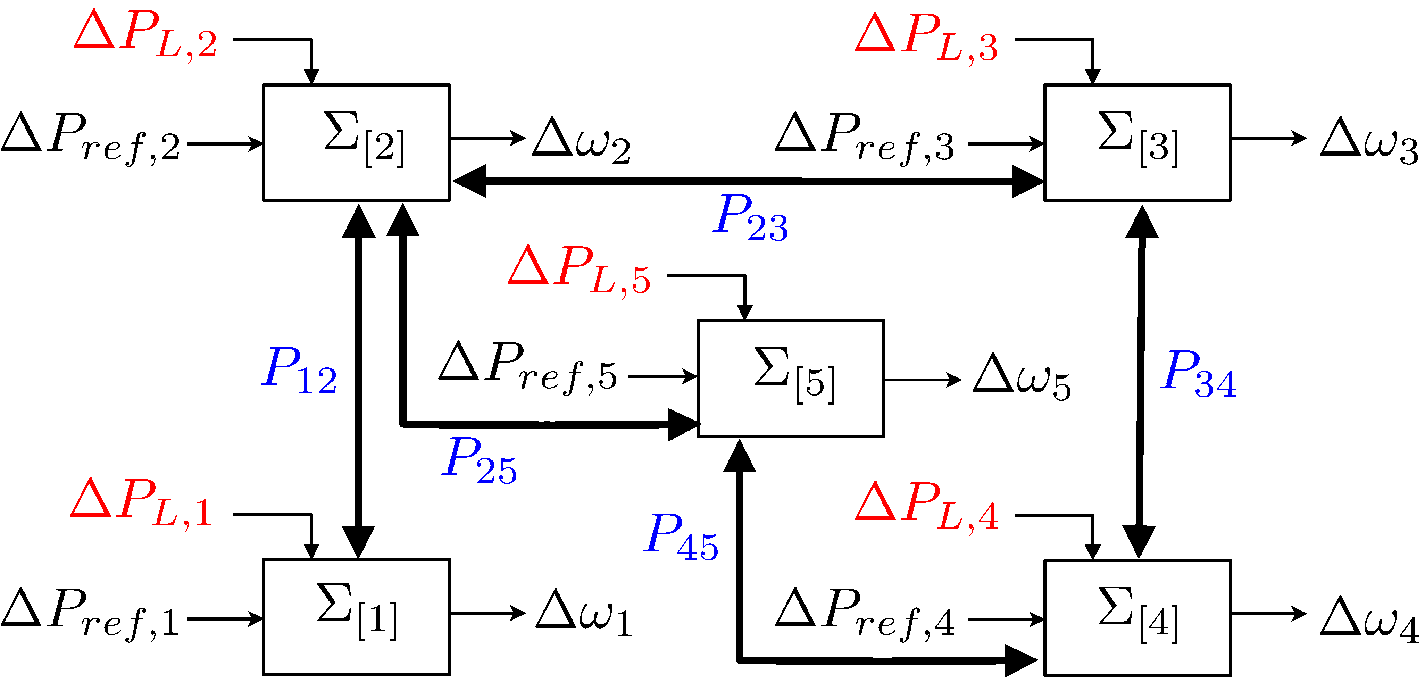}
                 \caption{Power network system of Scenario 2}
                 \label{fig:scenario2}
               \end{figure}
               
               As described in Section \ref{sec:plugin}, only systems $\subss\Sigma j$, $j\in\SSS_5$ update their controller $\subss\CC j$. For systems $\subss\Sigma j$, $j\in\SSS_5$, since the set $\NN_j$ changes, we retune controllers $\subss\CC j$ using Algorithm \ref{alg:distrisynt}. In particular, we compute $K_j$, $j\in\SSS_5$ and $K_5$ using the procedure described in Section \ref{sec:autoKi} with $\mu_{\alpha_k}=1$ and $\mu_{\beta_k}=1$, $k\in\{5\}\bigcup\SSS_5$ and obtain
               \begin{equation}
                 \label{eq:controllersKscen2}
                 \oneblock{      
                   K_2 &= -\matr{ 0.659 & 1.275 & 0.028 & 0.007 },~K_4 &= -\matr{ 0.713 & 1.105 & 0.048 & 0.008 },\\
                   K_5 &= -\matr{ 0.123 & 0.158 & 0.007 & 0.001 },
                 }
               \end{equation}          
               that allow inequalities \eqref{eq:pseudoinequalities} to be verified for systems $\subss\Sigma j$, $j\in\SSS_5$ and $\subss\Sigma 5$. Therefore $\mbf K$ fulfills Assumption \ref{ass:stability}-(\ref{enu:fschur}). Setting $\delta_j=10^{-4}$, $j\in\SSS_5$ and $\delta_5=10^{-4}$, the execution of Algorithm \ref{alg:distrisynt} does not stop before completion and hence we compute the new sets $\Zset_j(\delta_j)$, $\hXset_j$ and $\Vset_j$, $j\in\{5\}\bigcup\SSS_5$. We highlight that no retuning of controllers $\subss\CC 1$ and $\subss\CC 3$ is needed since systems $\subss\Sigma 1$ and $\subss\Sigma 3$ are not neighbors to system $\subss\Sigma 5$.\\
               \begin{figure}[!ht]
                 \centering
                 \subfigure[\label{fig:simulationscen2freq}Frequency deviation in each area controlled by the proposed De-MPC (bold line) and centralized MPC (dashed line).]{
                   \includegraphics[scale=0.5]{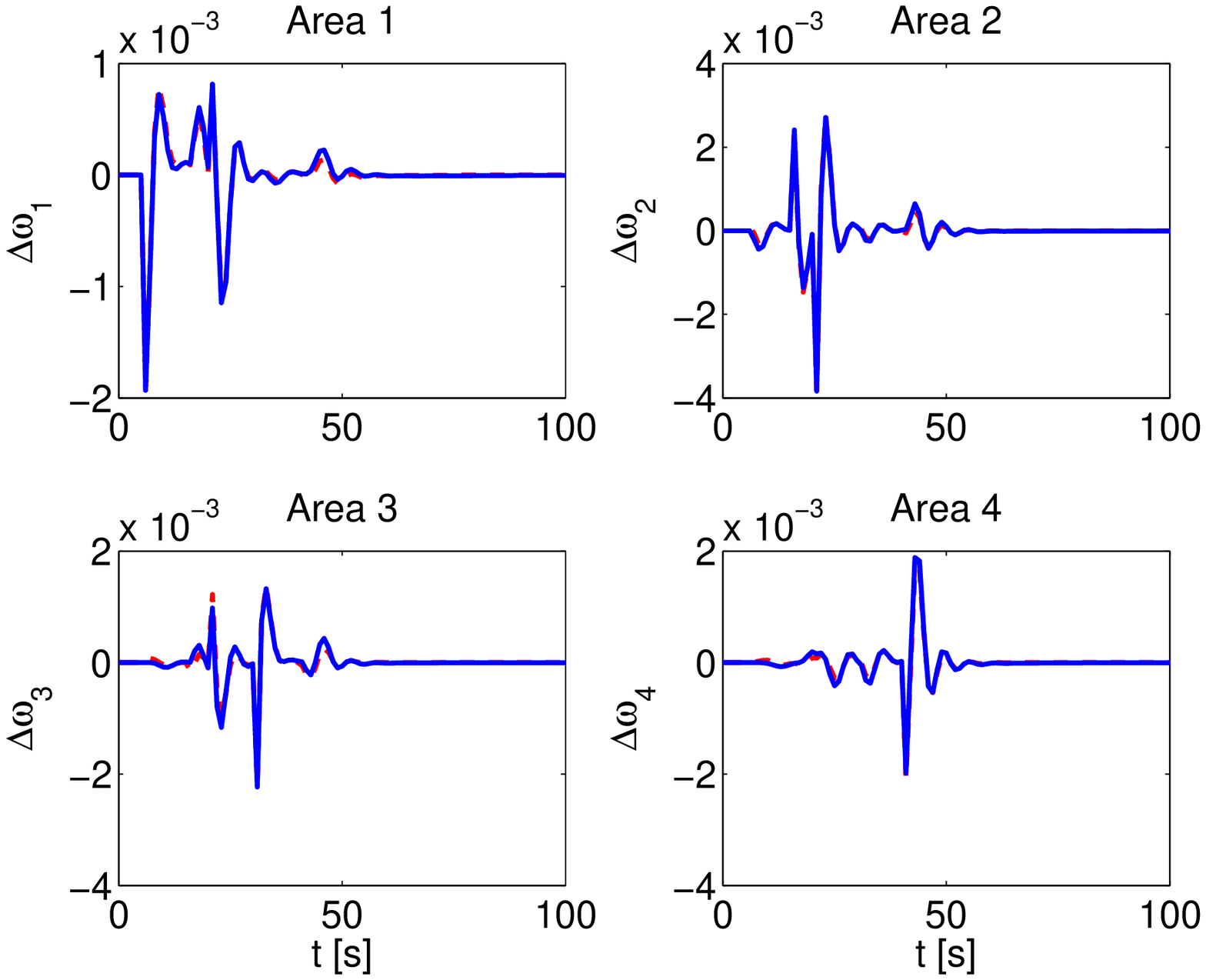}
                   \includegraphics[scale=0.475,width=4.5cm]{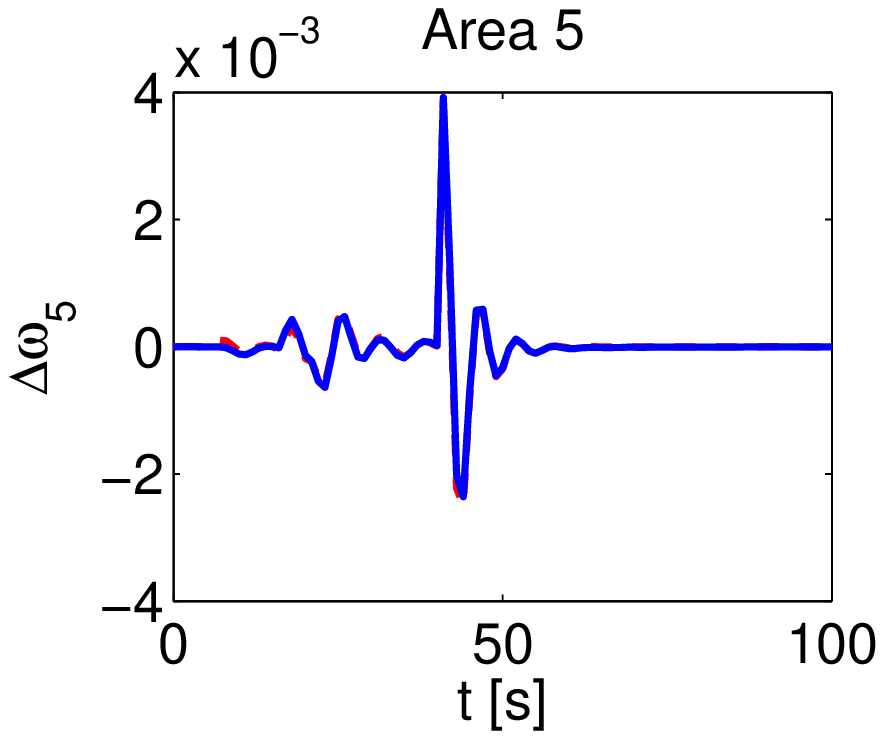}
                 }\\
                 \subfigure[\label{fig:simulationscen2ref}Load reference set-point in each area controlled by the proposed De-MPC (bold line) and centralized MPC (dashed line).]{
                   \includegraphics[scale=0.5]{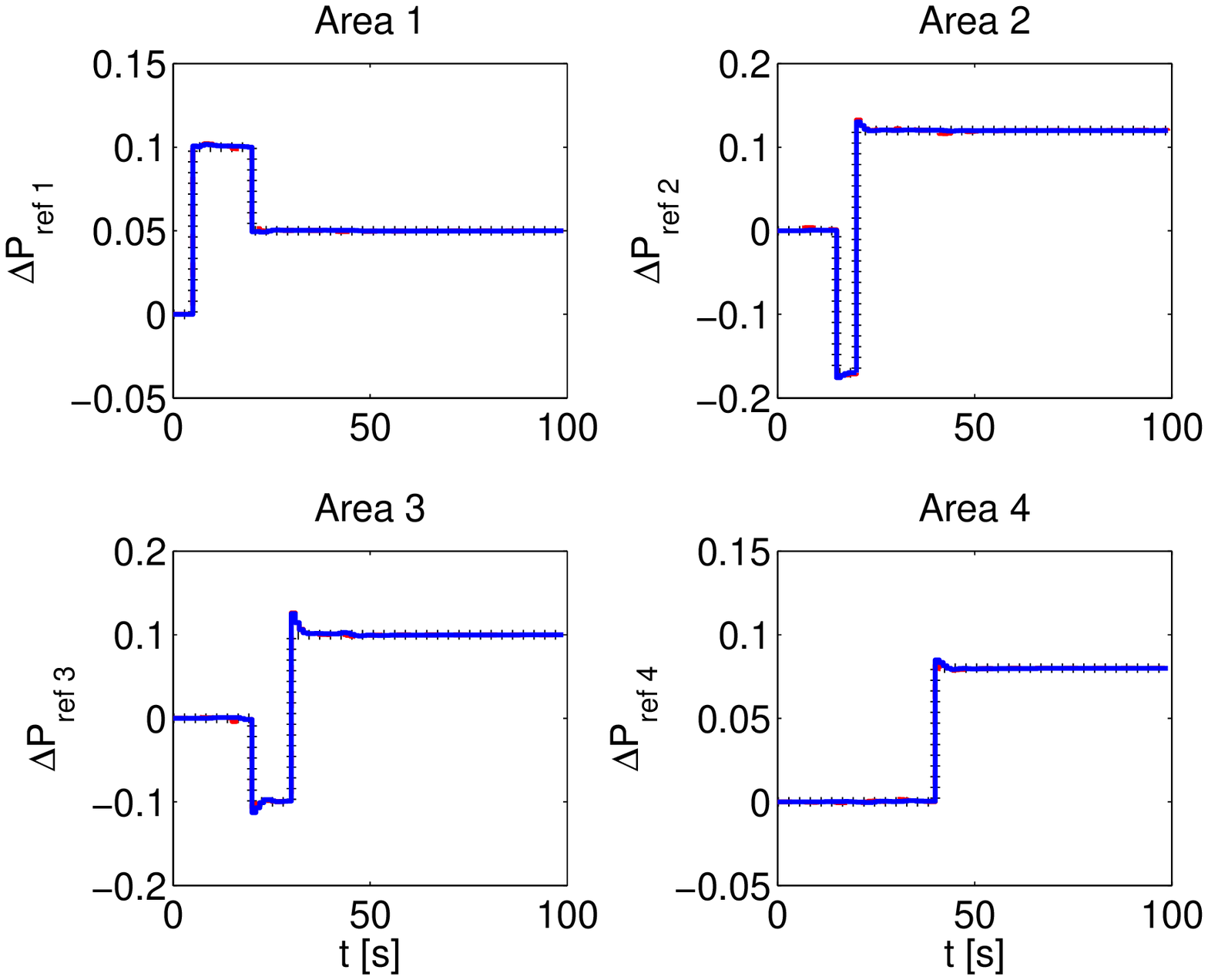}
                   \includegraphics[width=4.5cm,height=3.8cm]{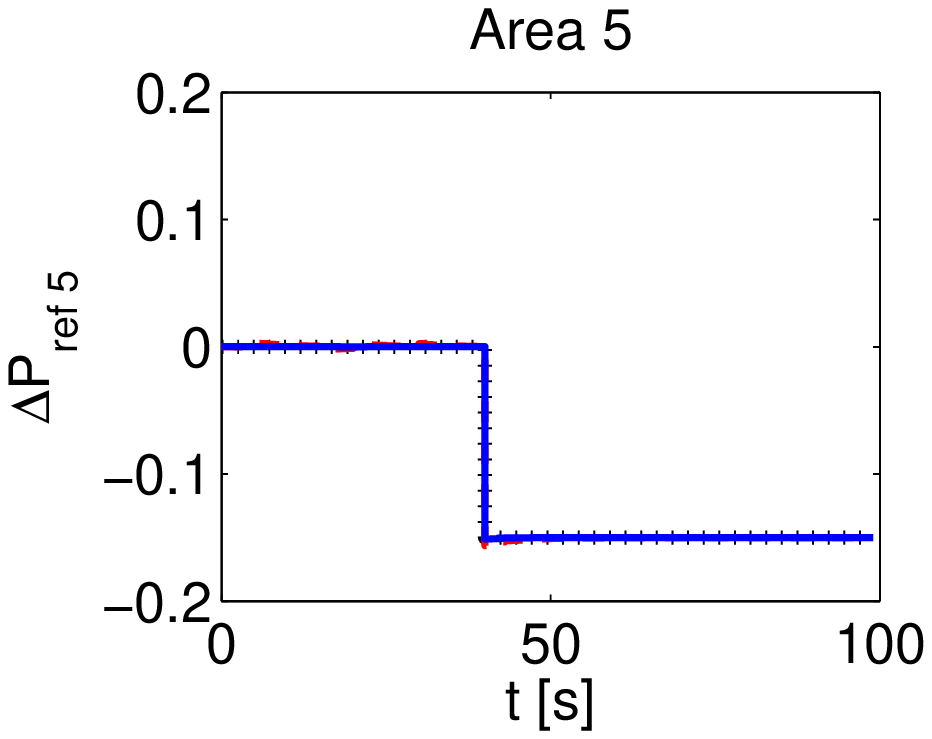}
                 }
                 \caption{Simulation Scenario 2: \ref{fig:simulationscen2freq} Frequency deviation and \ref{fig:simulationscen2ref} Load reference in each area.}
                 \label{fig:simulationscen2}
               \end{figure}
               \begin{figure}[!ht]
                 \centering
                 \includegraphics[scale=0.5]{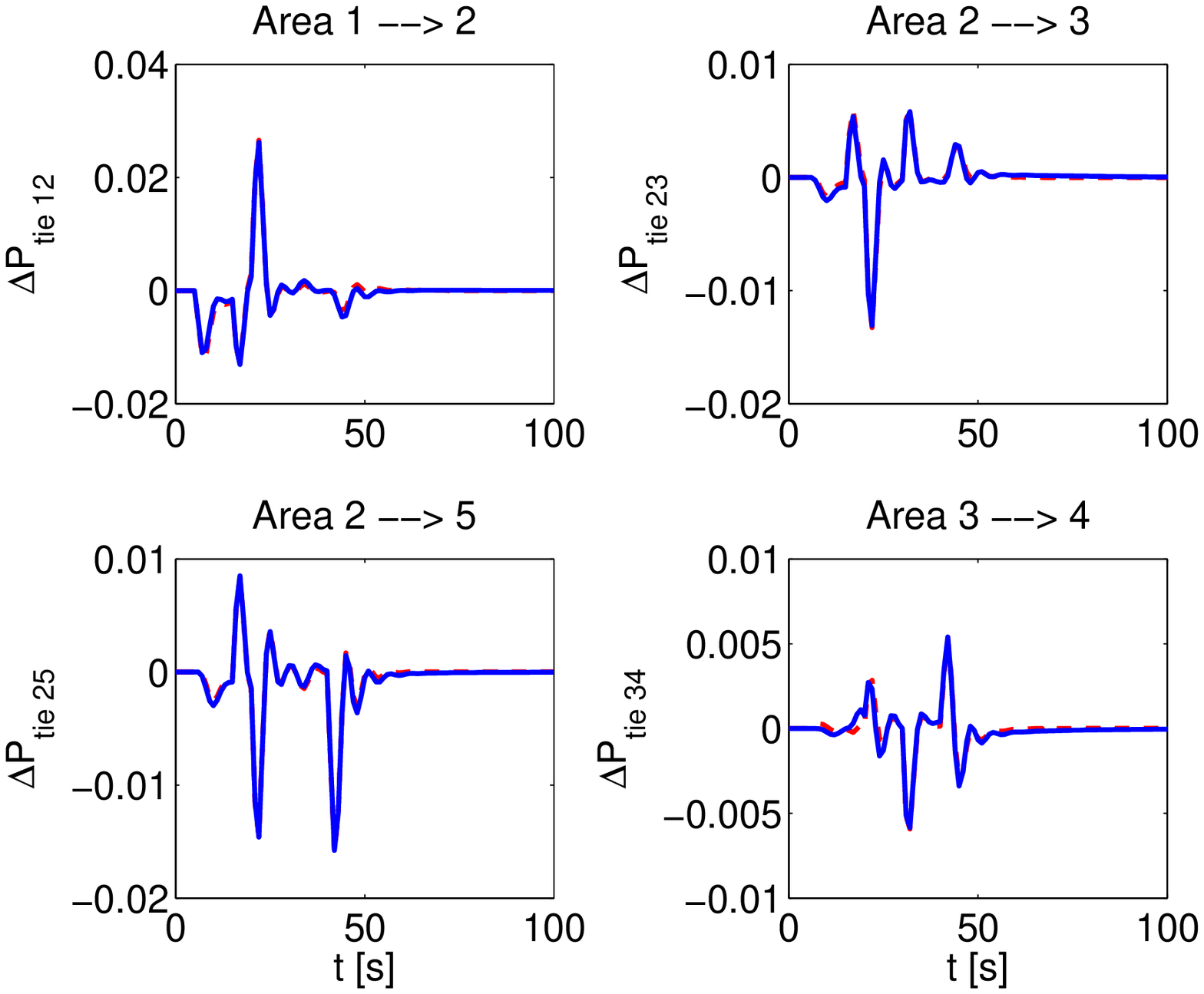}
                 \includegraphics[width=4.25cm,height=3.7cm]{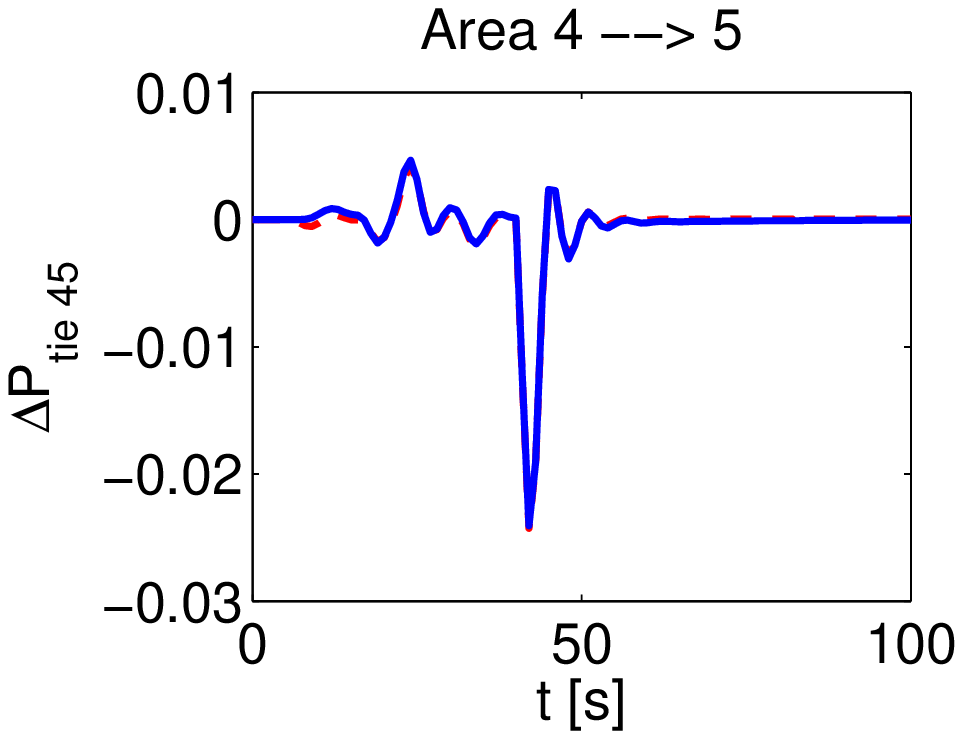}
                 \caption{Simulation Scenario 2: tie-line power between each area controlled by the proposed DeMPC (bold line) and centralized MPC (dashed line).}
                 \label{fig:simulationscen2tiepower}
               \end{figure}

               In Figure~\ref{fig:simulationscen2} we compare the performance of proposed DeMPC with the performance of centralized MPC. For centralized MPC we consider the overall system composed by the four areas, use the cost function $\sum_{i\in\MM}J_i^N$ and impose the collective constraints \eqref{eq:constraints}. The prediction horizon is $N_i=20,~i\in\MM$ for MPC-$i$ controllers and $N=20$ for centralized MPC. In the control experiment, step power loads $\Delta P_{L_i}$ specified in Appendix \ref{sec:valexe} have been used and they account for the step-like changes of the control variables in Figure \ref{fig:simulationscen2}. We highlight that the performance of decentralized and centralized MPC are totally comparable, in terms of frequency deviation (Figure~\ref{fig:simulationscen2freq}), control variables (Figure~\ref{fig:simulationscen2ref}) and power transfers $\Delta P_{{tie}_{ij}}$ (Figure \ref{fig:simulationscen2tiepower}).

          \subsection{Scenario 3}
               \label{sec:scenario3}
\begin{figure}[!ht]
                 \centering
                 \includegraphics[scale=0.75]{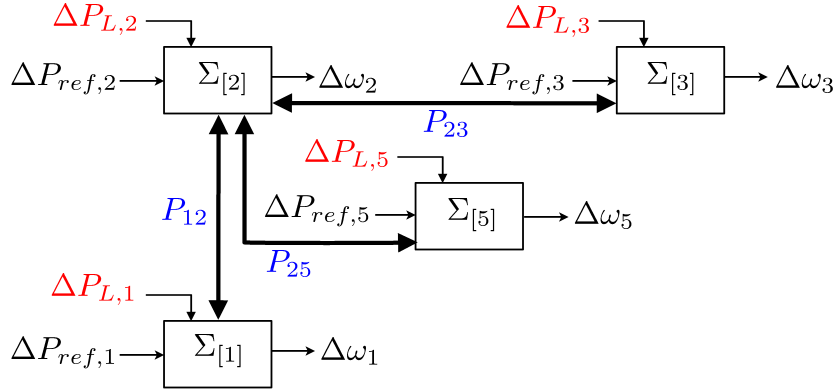}
                 \caption{Power network system of Scenario 3}
                 \label{fig:scenario3}
               \end{figure}

               We consider the power network described in Scenario 2 and disconnect the area $4$, hence obtaining the areas connected as in Figure \ref{fig:scenario3}. The set of successors to system 4 is $\SSS_4=\{3,5\}$.
               Because of disconnection, systems $\subss\Sigma j^C$, $j\in\SSS_4$ change their neighbors and local dynamics $A_{jj}$. Moreover, it is possible to verify that matrices $K_j$ computed in Scenario 2 do not solve Problem $\PP_j$, $j\in\SSS_4$. Then as described in Section \ref{sec:parametersdep}, each subsystem $\subss\Sigma j^C$, $j\in\SSS_4$ must retune controller $\subss\CC j$ by running Algorithm \ref{alg:distrisynt}. In particular, we compute $K_3$ and $K_5$ using the procedure proposed in Section \ref{sec:autoKi} with $\mu_{\alpha_j}=1$ and $\mu_{\beta_j}=1$, $j\in\SSS_4$ and obtain
               \begin{equation}
                 \label{eq:controllersKscen3}
                 \oneblock{      
                   K_3 &= -\matr{ 4.766 & 5.954 & 0.110 & 0.036 },~K_5 &= -\matr{ 4.102 & 4.861 & 0.201 & 0.038 }, \\
                 }
               \end{equation}
               that allows one to verify inequalities \eqref{eq:pseudoinequalities} for systems $\subss\Sigma j$, $j\in\SSS_4$. Therefore $\mbf K$ is such that Assumption \ref{ass:stability}-(\ref{enu:fschur}) holds. Setting $\delta_j=10^{-4}$, $j\in\SSS_4$, the execution of Algorithm \ref{alg:distrisynt} does not stop before completion and hence we compute the new sets $\Zset_j(\delta_j)$, $\hXset_j$ and $\Vset_j$, $j\in\SSS_4$. We highlight that retuning of controllers $\subss\CC 1$ and $\subss\CC 2$ is not needed since systems $\subss\Sigma 1$ and $\subss\Sigma 2$ are not neighbors to system $\subss\Sigma 4$.\\               
               \begin{figure}[!ht]
                 \centering
                 \subfigure[\label{fig:simulationscen3freq}Frequency deviation in each area controlled by the proposed De-MPC (bold line) and centralized MPC (dashed line).]{\includegraphics[scale=0.5]{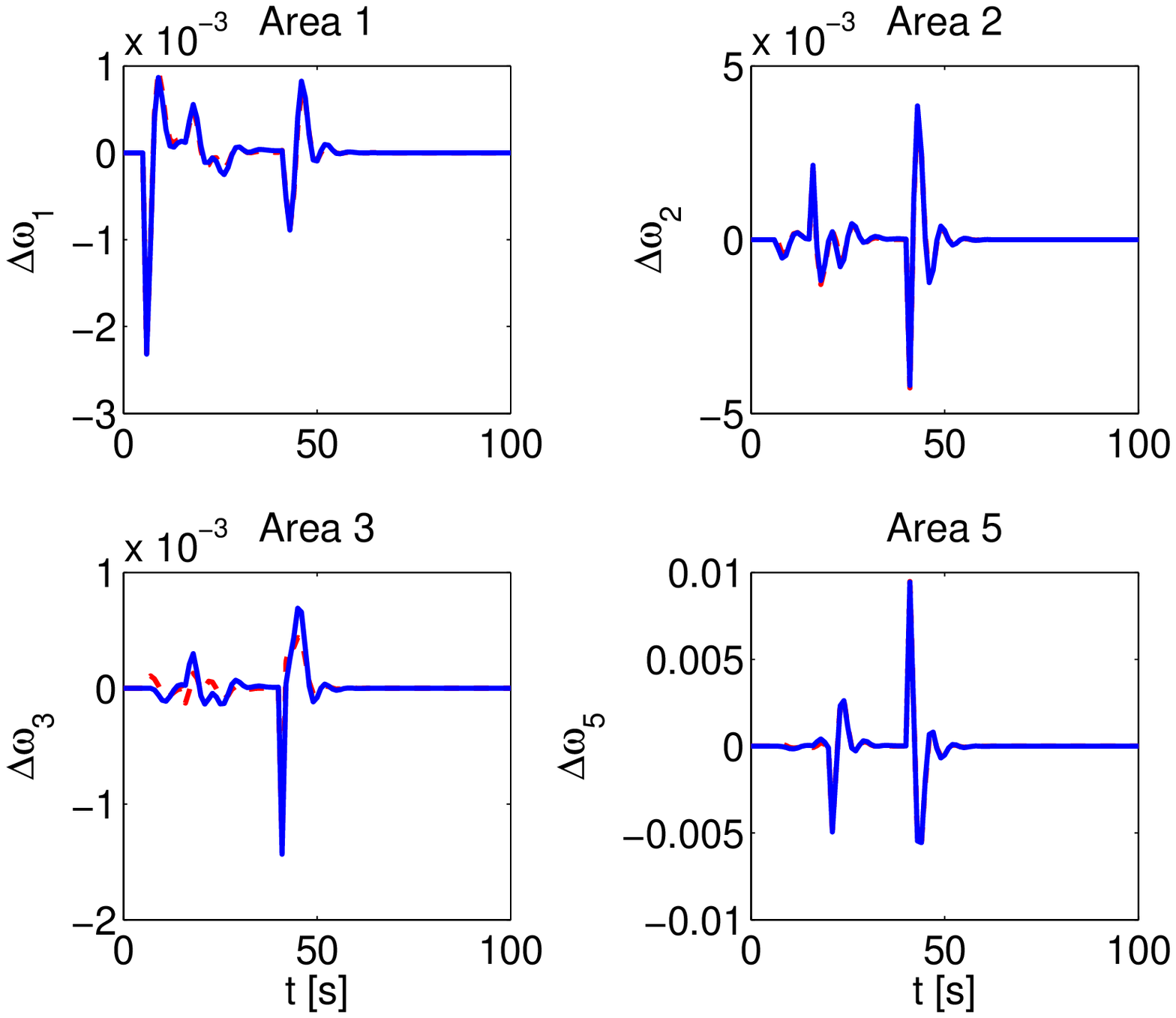}}\\
                 \subfigure[\label{fig:simulationscen3ref}Load reference set-point in each area controlled by the proposed De-MPC (bold line) and centralized MPC (dashed line).]{\includegraphics[scale=0.5]{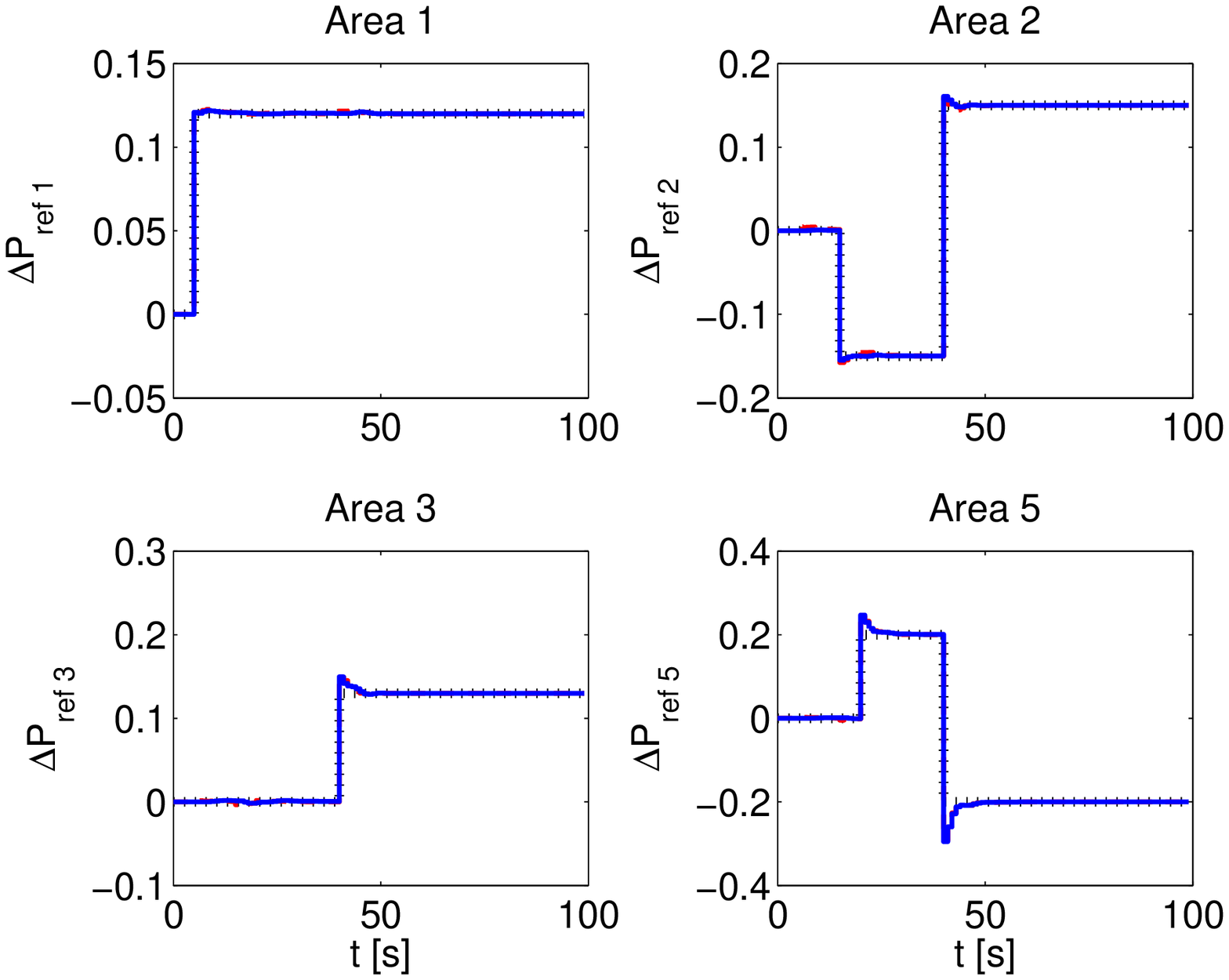}}
                 \caption{Simulation Scenario 3: \ref{fig:simulationscen3freq} Frequency deviation and \ref{fig:simulationscen3ref} Load reference in each area.}
                 \label{fig:simulationscen3}
               \end{figure}
               \begin{figure}[!ht]
                 \centering
                 \includegraphics[scale=0.5]{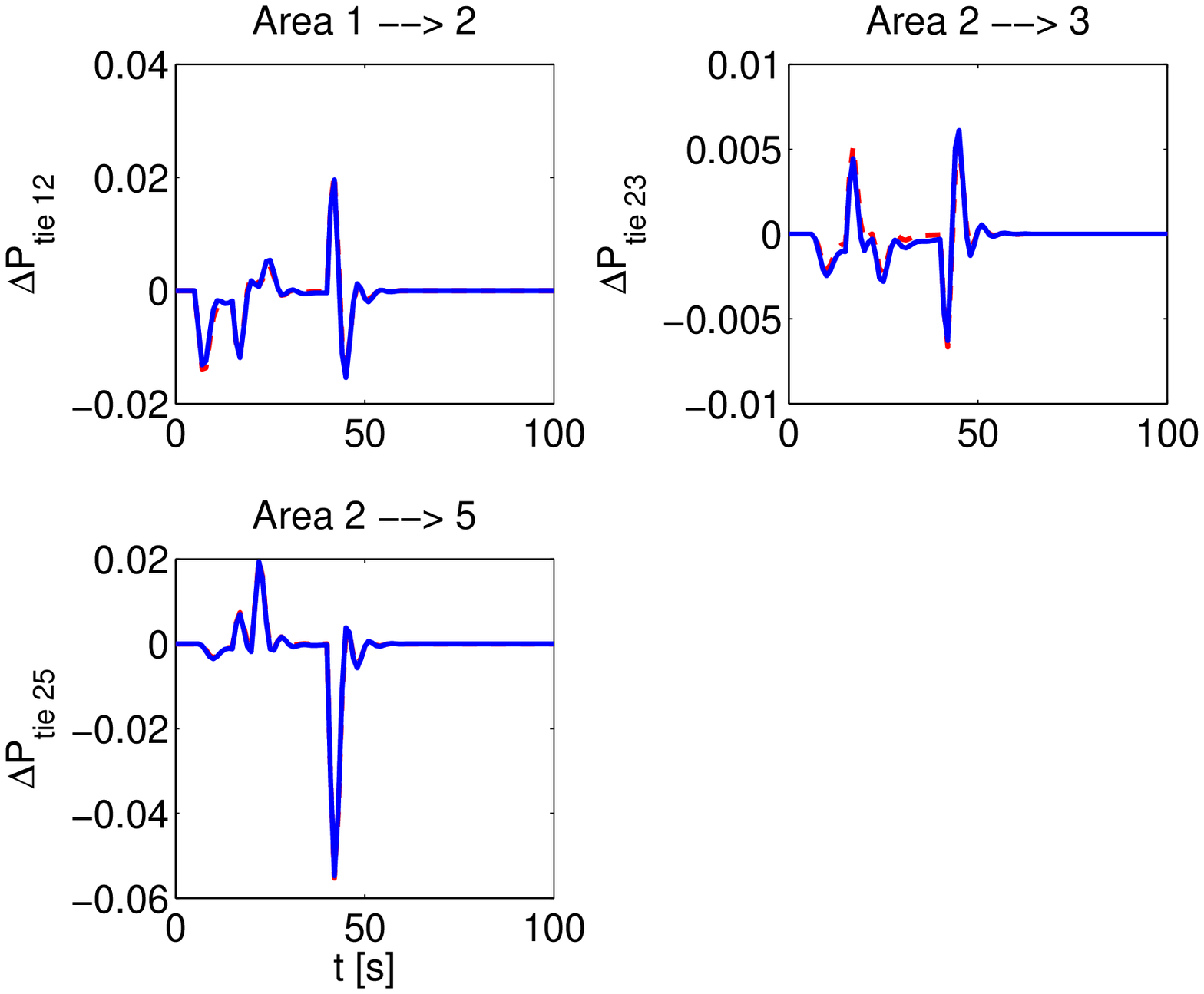}
                 \caption{Simulation Scenario 3: tie-line power between each area controlled by the proposed DeMPC (bold line) and centralized MPC (dashed line).}
                 \label{fig:simulationscen3tiepower}
               \end{figure}

               In Figure~\ref{fig:simulationscen3} we compare the performance of proposed DeMPC with the performance of centralized MPC. For centralized MPC we consider the overall system composed by the four areas, use the cost function $\sum_{i\in\MM}J_i^N$ and impose the collective constraints \eqref{eq:constraints}. The prediction horizon is $N_i=20,~i\in\MM$ for MPC-$i$ controllers and $N=20$ for centralized MPC. In the control experiment, step power loads $\Delta P_{L_i}$ specified in Appendix \ref{sec:valexe} have been used also in this case. We highlight that the performance of decentralized and centralized MPC are totally comparable in terms of frequency deviation (Figure~\ref{fig:simulationscen3freq}), control variables (Figure~\ref{fig:simulationscen3ref}) and power transfers $\Delta P_{{tie}_{ij}}$ (Figure \ref{fig:simulationscen3tiepower}).

     \section{Conclusions}
          \label{sec:conclusions}
          In this paper we proposed a tube-based DeMPC scheme for linear constrained systems, with the goal of stabilizing the origin of the closed-loop system and guaranteeing constraints satisfaction. The key feature of our approach is that the design procedure does not require any centralized computation. This enables P\&P operations, i.e. when a subsystem is plugged-in or unplugged at most the synthesis of its controller and the redesign of successors' controllers are needed. In future we will generalize our approach to embrace decentralized output-feedback control and tracking problems.

     \bibliographystyle{alpha}
     \bibliography{PnP_De_MPC_report}

     \appendix
          \section{Proof of Theorem \ref{thm:stabcon}}
               \label{sec:proofstabcon}
               The proof uses arguments similar to the ones adopted in \cite{Farina2012} for proving Theorem 1.
                 
               We first show recursive feasibility, i.e. that $\subss x i(t)\in\Xset_i^N,~\forall i\in\MM$ implies $\subss x i(t+1)\in\Xset_i^N$.
                 
               Assume that, at istant $t$, $\subss x i(t)\in\Xset_i^N$. The optimal nominal input and state sequences obtained by solving each MPC-$i$ problem $\Pset_i^N$ are $\subss v i(0:N_i-1|t)=\{\subss v i(0|t),\ldots,\subss v i(N_i-1|t)\}$ and $\subss \hx i(0:N_i|t)=\{\subss \hx i(0|t),\ldots,\subss \hx i(N_i|t)\}$, respectively. Define $\subss{v}{i}^{aux}(N_i|t)=\kappa_i^{aux}(\subss\hx i(N_i|t))$ and compute $\subss{\hx}{i}^{aux}(N_i+1|t)$ according to \eqref{eq:dynproblem} from $\subss\hx i(N_i|t)$ and $\subss v i(N_i|t)=\subss{v}{i}^{aux}(N_i|t)$. Note that, in view of constraint \eqref{eq:inTerminalSet} and points (\ref{enu:invariantAux}) and (\ref{enu:terminalSetAux}) of Assumption \ref{ass:axiomsMPC}, $\subss{v}{i}^{aux}(N_i|t)\in\Vset_i$ and  $\subss{\hx}{i}^{aux}(N_i+1|t)\in\hXset_{f_i}\subseteq\hXset_i$. We also define the input sequence 
               \begin{equation}
                 \label{eq:feasibleInput}
                 \subss\bv i(1:N_i|t)=\{\subss v i(1|t),\ldots,\subss v i(N_i-1|t),\subss{v}{i}^{aux}(N_i|t)\}
               \end{equation}
               and the state sequence produced by \eqref{eq:dynproblem} from the initial condition $\subss\hx i(0|t)$ and the input sequence $\subss\bv i(1:N_i|t)$, i.e. 
               \begin{equation}
                 \label{eq:feasibleState}
                 \subss\hbx i(1:N_i+1|t+1)=\{\subss\hx i(1|t),\ldots,\subss\hx i(N_i|t),\subss{\hx}{i}^{aux}(N_i+1|t)\}.
               \end{equation}
               In view of the constraints \eqref{eq:decMPCProblem} at time $t$ and recalling that $\Zset_i(\delta_i)$ is an RPI for \eqref{eq:decMPCProblem} and $\subss w i\in\Wset_i=\bigoplus_{j\in\NN_i}A_{ij}\Xset_j$, we have that $\subss x i(t+1)-\subss\hx i(1|t)\in\Zset_i(\delta_i)$. Therefore, we can conclude that the state and the input sequences $\subss\hbx i(1:N_i+1|t)$ and $\subss\bv i(1:N_i|t)$ are feasible at $t+1$, since constraints \eqref{eq:inZproblem}-\eqref{eq:inTerminalSet} are satisfied. This proves recursive feasibility.   
             
               We now prove convergence of the optimal cost function.

               We define $\Pset_i^{N,0}(\subss\hx i(0|t))=\min_{\subss v i(0:N_i-1|t)}\sum_{k=0}^{N_i-1}\ell_i(\subss\hx i(k),\subss v i(k))+V_{f_i}(\subss\hx i(N_i))$ subject to the constraints \eqref{eq:dynproblem}-\eqref{eq:inTerminalSet}. By optimality, using the feasible control law \eqref{eq:feasibleInput} and the corresponding state sequence \eqref{eq:feasibleState} one has
               \begin{equation}
                 \label{eq:optimcost}
                 \Pset_i^{N,0}(\subss\hx i(1|t))\leq\sum_{k=1}^{N_i}\ell_i(\subss\hx i(k|t),\subss v i(k|t))+V_{f_i}(\subss{\hx}{i}^{aux}(N_i+1|t+1))
               \end{equation}
               where it has been set $\subss v i(N_i|t)=\subss{v}{i}^{aux}(N_i|t)$. Therefore we have
               \begin{equation}
                 \label{eq:decrease}
                 \begin{aligned}
                   \Pset_i^{N,0}(\subss\hx i(1|t))-\Pset_i^{N,0}(\subss\hx i(0|t))\leq&-\ell_i(\subss\hx i(0|t),\subss v i(0|t))+\ell_i(\subss\hx i(N_i|t),\subss{v}{i}^{aux}(N_i|t))+\\
                   &+V_{f_i}(\subss{\hx}{i}^{aux}(N_i+1|t))-V_{f_i}(\subss{\hx}{i}^{aux}(N_i|t)).
                 \end{aligned}
               \end{equation}
               In view of Assumption \ref{ass:axiomsMPC}-(\ref{enu:decterminal}), from \eqref{eq:decrease} we obtain
               \begin{equation}
                 \label{eq:decreasLyap}
                 \Pset_i^{N,0}(\subss\hx i(1|t))-\Pset_i^{N,0}(\subss\hx i(t))\leq-\ell_i(\subss\hx i(t),\subss v i(t))
               \end{equation}
               and therefore $\subss\hx i(0|t)\rightarrow 0$ and $\subss v i(0|t)\rightarrow 0$ as $t\rightarrow\infty$.

                 Next we prove convergence to zero of state trajectories $\mbf x(t)$ of the closed-loop system with $\mbf x(0)\in\Xset^N$.

                 Recall that the state $\mbf x(t)$ evolves according to the equation \eqref{eq:controlled_model}. By asymptotic convergence to zero of the nominal state and input signals $\subss\hx i(0|t)$ and $\subss v i(0|t)$ respectively, using the diagonal structure of $\mbf B$ and $\mbf K$, we obtain that $\mbf B(\mbf v(0|t)-\mbf K\mbf\hx(0|t))$ is an asymptotically vanishing term. Under Assumption \ref{ass:stability}-(\ref{enu:fschur}), $\mbf{A+BK}$ is Schur, hence we obtain $\mbf x(t)\rightarrow 0$ as $t\rightarrow\infty$.

                 We prove now stability of the origin of the closed-loop system therefore completing the proof of statement (ii). We first show that
                 \begin{equation}
                   \label{eq:prop2mayne}
                   \mbf x(0)\in\Zset\Rightarrow\mbf x(t)\in\Zset\mbox{ and }\mbf x(t+1)=(\mbf{A+BK})\mbf x(t),~\forall t\geq 0
                 \end{equation}
                 where $\Zset=\prod_{i\in\MM}\Zset_i(\delta_i)$. Formula \eqref{eq:prop2mayne} is an easy consequence of Proposition 2 in \cite{Mayne2005} that we detail in the following for the sake of completeness.\\
                 If $\mbf x(0)\in\Zset$ then, as shown in \cite{Mayne2005}, $\mbf v=0$ and $\mbf \bx=0$. Therefore from \eqref{eq:controlled_model} we have
                 $$
                 \mbf\px(1)=(\mbf{A+BK})\mbf x(0)
                 $$
                 Furthermore, since sets $\Zset_i$ are RPI for \eqref{eq:errorDyn}, one has that $\Zset$ is positively invariant for
                 \begin{equation}
                   \label{eq:dyninZoproof}
                   \mbf\px=(\mbf{A+BK})\mbf x
                 \end{equation}
                 that coincides with \eqref{eq:errorDyn} after renaming veriables $\subss z i$ as $\subss x i$. Therefore $\mbf x(t)\in\Zset$ for $t=1$ and, applying the previous argument recursively for $t\geq 1$, we have shown that \eqref{eq:prop2mayne} holds.\\
                 Now, we focus on stability. Given $\epsilon>0$, choose $\eta\in(0,1)$ such that
                 \begin{equation}
                   \label{eq:etaZ}
                   \eta\Zset\subset\ball{\epsilon}(0).
                 \end{equation}
                 Such an $\eta$ always exists because $\Zset$ is bounded and includes the origin in its interior. More precisely, boundedness of $\Zset$ follows from $\Zset\subset\Xset$ and the boundedness of $\Xset$, that is guaranteed by Assumption \ref{ass:shapesets}. Furthermore, the mRPI for \eqref{eq:errorDyn} is given by \cite{Rawlings2009}
                 $$
                 \underline\Zset_i=\bigoplus_{k=0}^\infty F_i^k\Wset_i
                 $$
                 and therefore it includes the origin in its interior. It follows that the same is true for sets $\Zset_i(\delta_i)$, $i\in\MM$ and $\Zset=\prod_{i\in\MM}\Zset_i(\delta_i)$. Since the origin is strictly contained in $\Zset$, there always exists $\underline\delta>0$ such that $\ball{\underline\delta}(0)\subseteq\eta\Zset$. Since $\ball{\underline\delta}(0)\subset\Zset$ one has that, in view of \eqref{eq:prop2mayne}, the state trajectory $\mbf x(t),~t\geq 0$ stemming from $\mbf x(0)\in\ball{\underline\delta}(0)$ fulfill the dynamics \eqref{eq:dyninZoproof}. Furthermore, since \eqref{eq:dyninZoproof} is a linear system for which $\Zset$ is positively invariant set, one has that also $\eta\Zset$ is positively invariant. Then, we have shown that
                 $$
                 \mbf x(0)\in\ball{\underline\delta}(0)\rightarrow \mbf x(t)\in\eta\Zset,~\forall t\geq 0.
                 $$
                 From \eqref{eq:etaZ} stability of the origin follows.
                 \begin{flushright}$\blacksquare$\end{flushright}

          \section{Proof of Theorem \ref{thm:main}}
               \label{sec:proofmain}
                    \subsection{Proof of (I)}
                    Define a matrix $\Metz$ such that its $ij$-th entry $\mu_{ij}$ is
                    $$
                    \begin{array}{lcl}
                      \mu_{ij}=-1&\text{if}&i=j\\
                      \mu_{ij}=\sum_{k=0}^{\infty}\|\mathcal{F}_{i}F_{i}^kA_{ij}
                      \mathcal{F}_{j}^{\flat}\|_{\infty}&\text{if}&i\neq j.
                    \end{array}
                    $$
Note that all the off-diagonal entries of matrix $\Metz$ are non-negative, i.e., it is Metzler~\cite{Farina2000}. We recall the following results.
                    \begin{lem}[see \cite{Mason2007}]
                      \label{lem:metzhur}
                      Let matrix $\Metz\in\Rset^{M\times M}$ be Metzler. Then $\Metz$ is Hurwitz if and only if there is a vector $\nu\in\Rset_+^M$ such that $\Metz\nu<\Zero$.
                    \end{lem}
                    \begin{lem}[]
                      \label{lem:metzschur}
                      Define the matrix $\Gamma=\Metz+\eye M$
                      where $\Metz\in\Rset^{M\times M}$, $\eye M$ is the $M\times M$ identity matrix and  $\Gamma$ is non negative. Then the Metzler matrix $\Metz$ is Hurwitz if and only if $\Gamma$ is Schur.
                    \end{lem}
                    The proof of Lemma~\ref{lem:metzschur} easily follows from Theorem 13 in \cite{Farina2000}.\\
                    Inequalities~\eqref{eq:pseudoinequalities} are equivalent to $\Metz\nu<\Zero_M$ where $\nu=\One_M$. Then, from Lemma \ref{lem:metzhur}, $\Metz$ is Hurwitz. From Lemma \ref{lem:metzschur}, \eqref{eq:pseudoinequalities} implies that matrix $\Gamma= \Metz+\eye M$ is Schur.\\
                    For system $\subss\Sigma i$ in \eqref{eq:subsystem}-\eqref{eq:couplingW}, when $\subss u i$ is defined as in \eqref{eq:tubecontrol}, $\subss v i=0$ and $\subss\bx i=0$, we have
                    \begin{equation}
                      \label{eq:lagruzero}
                      \subss x i (t)=F_i^t\subss x i(0)+\sum_{k=0}^{t-1}F_i^k\sum_{j\in\NN_i}A_{ij}\subss x j(t-k-1)
                    \end{equation}
                    In view of \eqref{eq:lagruzero} we can write
                    \begin{equation*}
                      \label{eq:lagruzeronorm}
                      \oneblock{
                        \norme{\FF_i\subss x i (t)}{\infty}
                                                            &\leq\norme{\FF_iF_i^t\FF_i^\flat}{\infty}\norme{\FF_i\subss x i(0)}{\infty}+\\ &+\sum_{j\in\mathcal{N}_i}\gamma_{ij}\max_{\substack{k\leq t}}\norme{\FF_j \subss x j(k)}{\infty}.
                      }
                    \end{equation*}
where $\gamma_{ij}$ are the entries of $\Gamma$. Denoting $\subss \tx i=\FF_i \subss x i$, we can collectively define $\tilde{\mbf x}=\tilde{\mbf\FF}\mbf x$, where $\tilde{\mbf\FF}=\diag(\FF_1,\dots,\FF_M)$. From the definition of sets $\Xset_i$, we have rank$(\tilde{\mbf\FF})=n$. We define the system
                    \begin{align}
                      \label{eq:exp_sys}
                      \mbf \tx^+=(\tilde{\mbf A}+\tilde{\mbf B}\tilde{\mbf K})\tilde{\mbf x}
                    \end{align}
                    where $\tilde{\mbf A}=\tilde{\mbf\FF}\mbf A\tilde{\mbf\FF}^{\flat}$, $\tilde{\mbf B}=\tilde{\mbf\FF}\mbf B$ and $\tilde{\mbf K}=\mbf K\tilde{\mbf\FF}^{\flat}$. In order to analyze the stability of the origin of \eqref{eq:exp_sys}, we consider the method proposed in \cite{Dashkovskiy2007}.
                    In view of Corollary 16 in \cite{Dashkovskiy2007}, the overall system \eqref{eq:exp_sys} is asymptotically stable if the gain matrix $\Gamma$ is Schur. As shown above this property is implied by \eqref{eq:pseudoinequalities}.\\
Moreover, system~\eqref{eq:exp_sys} is an expansion of the original system (see Chapter 3.4 in \cite{Lunze1992}). In view of the inclusion principle (see Theorem 3.3 in \cite{Lunze1992} and \cite{Stankovic2004} for a discrete-time version), the asymptotic stability of \eqref{eq:exp_sys} implies the asymptotic stability of the original system.
               \subsection{Proof of (II)}
                    First note that, for $i\in\MM$, in view of~\eqref{eq:statezonotope} $\norme{f_{i,r}^T\Xi_i}{\infty}=1$ for all $r\in1:\bar{r}_i$ and therefore $\norme{\FF_i\Xi_i}{\infty}=1$. This implies that $\norme{f_{i,r}^TF_{i}^kA_{ij}\Xi_j}{\infty} \leq\norme{f_{i,r}^TF_{i}^kA_{ij}\mathcal{F}_{j}^{\flat}}{\infty}\norme{\mathcal{F}_{j}\Xi_j}{\infty}=\norme{f_{i,r}^TF_{i}^kA_{ij}\mathcal{F}_{j}^{\flat}}{\infty}
                    \leq\norme{\mathcal{F}_{i}F_{i}^kA_{ij}\mathcal{F}_{j}^{\flat}}{\infty}$. Therefore, in view of~\eqref{eq:pseudoinequalities}, for all $r\in 1:\bar{r}_i$
                    \begin{equation}
                      \label{eq:statecondition}
                      \sum_{k=0}^\infty\sum_{j\in\NN_i}\norme{f_{i,r}F_i^kA_{ij}\Xi_j}{\infty}\leq \sum_{k=0}^\infty\sum_{j\in\NN_i}\norme{\mathcal{F}_{i}F_{i}^kA_{ij}\mathcal{F}_{j}^{\flat}}{\infty} <1
                    \end{equation}
                    Now we want to find parameter $\hl_i>0$ such that, simultaneously, the inclusion \eqref{eq:stateinclu} holds and
                    $\Zset_i(\delta_i)$ is a $\delta_i-$outer approximation of the mRPI $\underline\Zset_i$.
                    The mRPI for \eqref{eq:errorDyn} is given by \cite{Rakovic2005a}
                    \begin{equation}
                      \label{eq:mRPI}
                      \underline\Zset_i=\bigoplus_{k=0}^\infty F_i^k\bigoplus_{j\in\NN_i}A_{ij}\Xset_j.
                    \end{equation}
                    From \cite{Rakovic2005a}, for given $\delta_i>0$ there exist $\alpha_i\in\Rset$ and $s_i\in\Nset_+$ such that the set
                    \begin{equation}
                      \label{eq:approxmRPI}
                      \Zset_i(\delta_i)=(1-\alpha_i)^{-1}\bigoplus_{k=0}^{s_i-1} F_i^k\bigoplus_{j\in\NN_i}A_{ij}\Xset_j
                    \end{equation}
                    is a $\delta_i-$outer approximation of the mRPI $\underline\Zset_i$.\\
                    Define $\bXset_i=\hXset_i\oplus\Zset_i(\delta_i)$. Following the proof of Proposition 2 in \cite{Farina2012} and using arguments from Section 3 of \cite{Kolmanovsky1998}, we can then guarantee \eqref{eq:stateinclu} if $\bXset_i\subseteq\Xset_i$, which holds if, for all $r\in 1:\bar{r}_i$
                    \begin{equation}
                      \label{eq:supZhX}
                      \sup_{\substack{\subss{z}{i}\in\Zset_i(\delta_i)\\\subss{\hx}{i}\in\hXset_i}}~f_{i,r}^T(\subss{z}{i}+\subss{\hx}{i})\leq 1.
                    \end{equation}
                    Using \eqref{eq:defapproxmRPI} and \eqref{eq:mRPI}, the inequalities \eqref{eq:supZhX} are verified if
                    \begin{equation}
                      \label{eq:supmZhX}
                      \sup_{\substack{\{\subss{x}{j}(k)\in \Xset_j\}_{j\in\NN_i}^{k=0,\ldots,\infty}\\\subss{\hx}{i}\in\hXset_i\\\sigma_i\in\ball{\delta_i}(0)}}
                      \!\!\!\!\!\!\!\!\!\!\!h^x_{i,r}(\{\subss{x}{j}(k)\}_{j\in\NN_i}^{k=0,\ldots,\infty},\subss{\hx}{i})+
                      \norme{f_{i,r}^T\sigma_i}{\infty}\leq 1
                    \end{equation}
                    where $h^x_{i,r}(\cdot)=f_{i,r}^T(\sum_{k=0}^\infty F_i^k\sum_{j\in\NN_i}A_{ij}\subss{x}{j}(k)+\subss{\hx}{i})$.\\
                    Since $\norme{f_{i,r}^T\sigma_i}{\infty}\leq\norme{f_{i,r}^T}{\infty}\delta_i$, conditions \eqref{eq:supmZhX} are satisfied if
                    \begin{equation}
                      \label{eq:supmZhXwB}
                      \sup_{\substack{\{\subss{x}{j}(k)\in \Xset_j\}_{j\in\NN_i}^{k=0,\ldots,\infty}\\\subss{\hx}{i}\in\hXset_i}}\!\!\!\!\!\!\!\!h^x_{i,r}(\{\subss{x}{j}(k)\}_{j\in\NN_i}^{k=0,\ldots,\infty},\subss{\hx}{i})\leq 1-\norme{f_{i,r}^T}{\infty}\delta_i.
                    \end{equation}
                    Using \eqref{eq:statezonotope} and \eqref{eq:tightenedstatezonotope} we can rewrite \eqref{eq:supmZhXwB} as
                    \begin{equation}
                      \label{eq:supmZhX2}
                      \sup_{\substack{\{\norme{d_j(k)}{\infty}\leq 1\}_{j\in\NN_i}^{k=0,\ldots,\infty}\\\norme{\hd_i}{\infty}\leq\hl_i}}
                      \!\!\!\!\!\!\!\!\!h^d_{i,r}(\{d_j(k)\}_{j\in\NN_i}^{k=0,\ldots,\infty},\hd_i)
                      \leq 1-\norme{f_{i,r}^T}{\infty}\delta_i
                    \end{equation}
                    where $h^d_{i,r}(\cdot)=f_{i,r}^T(\sum_{k=0}^\infty F_i^k\sum_{j\in\NN_i}A_{ij}\Xi_jd_j(k)+\hXi_i\hd_i)$.\\
                    The inequalities \eqref{eq:supmZhX2} are satisfied if
                    \begin{equation}
                      \label{eq:supnorm}
                      \sum_{k=0}^\infty\sum_{j\in\NN_i}\norme{f_{i,r}^TF_i^kA_{ij}\Xi_j}{\infty}+\norme{f_{i,r}^T\hXi_i}{\infty}\hl_i\leq 1-\norme{f_{i,r}^T}{\infty}\delta_i
                    \end{equation}
                    for all $r\in1:\bar{r}_i$.\\
                    In view of~\eqref{eq:statecondition} there exist sufficiently small $\delta_i>0$ and $\hl_i>0$ satisfying~\eqref{eq:supnorm} (and therefore verifying \eqref{eq:stateinclu}), e.g. choosing $\hl_i\in(0,{\hat L}_i]$.
               \subsection{Proof of (III)}

                    For each $i\in\MM$, we want to find tightened input constraint $\Vset_i$ such that \eqref{eq:inputinclu} holds. Following the rational used in Section 3 of \cite{Kolmanovsky1998}, from definition of sets $\Uset_i$ and $\Vset_i$, \eqref{eq:inputinclu} holds if \eqref{eq:inputinequalities} is satisfied. Hence, choosing $\Vset_i$ as in \eqref{eq:tightenedinputconstraints}, for $l_{v_{i,r}}=\hl_{v_{i,r}}(\delta_i)$ the inclusion \eqref{eq:inputinclu} holds.
                    \begin{flushright}$\blacksquare$\end{flushright}

          \section{Parameters, constraints and setpoints of experiment described in Section \ref{sec:example}}
               \label{sec:valexe}                 
               \begin{table}[!ht]
                 \footnotesize
                 \centering
                 \begin{tabular}{|c|c|}
                   \hline
                   $\Delta\theta_i$ & Deviation of the angular displacement of the rotor with respect to the stationary reference axis on the stator \\
                   $\Delta\omega_i$ & Speed deviation of rotating mass from nominal value\\
                   $\Delta P_{m_i}$ & Deviation of the mechanical power from nominal value (p.u.)\\
                   $\Delta P_{v_i}$ & Deviation of the steam valve position from nominal value (p.u.)\\
                   $\Delta P_{ref_i}$ & Deviation of the reference set power from nominal value (p.u.)\\
                   $\Delta P_{L_i}$ & Deviation of the nonfrequency-sensitive load change from nominal value (p.u.)\\
                   $H_i$ & Inertia constant defined as $H_i=\frac{\mbox{kinetic energy at rated speed}}{\mbox{machine rating}}$ (typically values in range $[1-10]\mbox{ sec}$) \\
                   $R_i$ & Speed regulation \\
                   $D_i$ & Defined as $\frac{\mbox{percent change in load}}{\mbox{change in frequency}}$ \\
                   $T_{t_i}$ & Prime mover time constant (typically values in range $[0.2-2]\mbox{ sec }$)\\
                   $T_{g_i}$ & Governor time constant (typically values in range $[0.1-0.6]\mbox{ sec }$) \\
                   $P_{ij}$ & Slope of the power angle curve at the initial operating angle between area $i$ and area $j$ \\
                   \hline
                 \end{tabular}
                 \caption{Variables of a generation area with typical value ranges \cite{Saadat2002}. (p.u.) stands for ``per unit''.}
                 \label{tab:networkparameter}
               \end{table}
               \normalsize 

               \begin{table}[!ht]
                 \centering
                 \begin{tabular}{|c|c|c|c|c|c|}
                   \hline
                   &  Area 1 & Area 2 & Area 3 & Area 4 & Area 5 \\
                   \hline
                   $H_i$     &  12       & 10        &  8        &  8        & 10         \\
                   \hline
                   $R_i$     &   0.05   & 0.0625 & 0.08    &  0.08   &  0.05    \\
                   \hline
                   $D_i$     &  0.7      & 0.9      & 0.9      &  0.7     &  0.86     \\
                   \hline
                   $T_{t_i}$ &   0.65    & 0.4      & 0.3      &  0.6     &  0.8       \\
                   \hline
                   $T_{g_i}$ &  0.1      & 0.1      & 0.1      &  0.1     &   0.15    \\
                   \hline
                   \multicolumn{6}{c}{}                                                          \\
                 \end{tabular}
                 
                 \begin{tabular}{|c|c|c|c|c|c|}
                   \hline
                   &  Area 1 & Area 2 & Area 3 & Area 4 & Area 5 \\
                   \hline
                   $\Delta\theta_i$    &  $\norme{\subss{x}{1,1}}{\infty}\leq 0.1$   &  $\norme{\subss{x}{2,1}}{\infty}\leq 0.1$   &  $\norme{\subss{x}{3,1}}{\infty}\leq 0.1$ &   $\norme{\subss{x} {4,1}}{\infty}\leq 0.1$   &  $\norme{\subss{x}{5,1}}{\infty}\leq 0.1$    \\
                   \hline
                   $\Delta P_{ref_i}$    &  $\norme{\subss{u}{1}}{\infty}\leq 0.5$   &  $\norme{\subss{u}{2}}{\infty}\leq 0.65$   &  $\norme{\subss{u}{3}}{\infty}\leq 0.65$ &   $\norme{\subss{u} {4}}{\infty}\leq 0.55$   &  $\norme{\subss{u}{5}}{\infty}\leq 0.5$    \\
                   \hline 
                   \multicolumn{6}{c}{}\\
                 \end{tabular}
                 
                 $P_{12} = 4\qquad P_{23}=2\qquad P_{34}=2\qquad P_{45}=3\qquad P_{25}=3$\\                
                 
                 \caption{Model parameters and constraints for systems $\subss\Sigma i,~i\in1:5$.}
                 \label{tab:scenario123}
               \end{table} 
               \begin{table}[!ht]
                 \centering
                 \begin{tabular}{|c|c|c|}
                   \hline
                   Step time  &  Area $i$ & $\Delta P_{L_i}$ \\
                   \hline
                   5               &      1        &   +0.15             \\
                   \hline
                   15             &      2        &   -0.15             \\
                   \hline
                   20             &      3        &   +0.12             \\
                   \hline
                   40             &      3        &   -0.12             \\
                   \hline
                   40             &      4        &   +0.28            \\
                   \hline
                 \end{tabular}
                 \caption{Load of power $\Delta P_{L_i}$ (p.u.) for simulation in Scenario 1. $+\Delta P_{L_i}$ means a step of required power, hence a decrease of the frequency deviation $\Delta\omega_i$ and then an increase of the power reference $\Delta P_{ref_i}$, while $-\Delta P_{L_i}$ means the opposite.}
                 \label{tab:simulationscen1}
               \end{table}

               \begin{table}[!ht]
                 \centering
                 \begin{tabular}{|c|c|c|}
                   \hline
                   Step time  &  Area $i$ & $\Delta P_{L_i}$ \\
                   \hline
                   5               &      1        &   +0.10             \\
                   \hline
                   15             &      2        &   -0.17             \\
                   \hline
                   20             &      1        &   +0.05             \\
                   \hline
                   20             &      2        &   +0.12             \\
                   \hline
                   20             &      3        &   -0.10             \\
                   \hline
                   30             &      3        &   +0.10             \\
                   \hline
                   40             &      4        &   +0.08             \\
                   \hline
                   40             &      5        &   -0.15             \\
                   \hline
                 \end{tabular}
                 \caption{Load of power $\Delta P_{L_i}$ (p.u.) for simulation in Scenario 2. $+\Delta P_{L_i}$ means a step of required power, hence a decrease of the frequency deviation $\Delta\omega_i$ and then an increase of the power reference $\Delta P_{ref_i}$, while $-\Delta P_{L_i}$ means the opposite.}
                 \label{tab:simulationscen2}
               \end{table}

               \begin{table}[!ht]
                 \centering
                 \begin{tabular}{|c|c|c|}
                   \hline
                   Step time  &  Area $i$ & $\Delta P_{L_i}$ \\
                   \hline
                   5               &      1        &   +0.12             \\
                   \hline
                   15             &      2        &   -0.15             \\
                   \hline
                   20             &      5        &   +0.20             \\
                   \hline
                   40             &      2        &   +0.15             \\
                   \hline
                   40             &      3        &   +0.13            \\
                   \hline
                   40             &      5        &   -0.20            \\
                   \hline
                 \end{tabular}
                 \caption{Load of power $\Delta P_{L_i}$ (p.u.) for simulation in Scenario 3. $+\Delta P_{L_i}$ means a step of required power, hence a decrease of the frequency deviation $\Delta\omega_i$ and then an increase of the power reference $\Delta P_{ref_i}$, while $-\Delta P_{L_i}$ means the opposite.}
                 \label{tab:simulationscen3}
               \end{table}

\end{document}